\documentclass[fleqn,usenatbib]{mnras}

\usepackage{newtxtext,newtxmath}

\usepackage[T1]{fontenc}

\DeclareRobustCommand{\VAN}[3]{#2}
\let\VANthebibliography\thebibliography
\def\thebibliography{\DeclareRobustCommand{\VAN}[3]{##3}\VANthebibliography}

\usepackage{graphicx}	
\usepackage{amsmath}	
\usepackage{comment}

\title[Type-C QPO of MAXI~J1348--630]{Variability and phase lags of the type-C quasi-periodic oscillation of MAXI J1348--630 with \textit{NICER}}

\author[K. Alabarta et al.]{
Kevin Alabarta,$^{1, 2}$\thanks{E-mail: k.alabarta@soton.ac.uk},
Mariano M\'endez$^{1}$, 
 Federico Garc\'ia$^{1, 3}$,
 Valentina Peirano$^{1}$,
 Diego Altamirano$^{2}$,
\newauthor
 Liang Zhang$^{2, 4}$ 
 and Konstantinos Karpouzas$^{1, 2}$
\\
$^{1}$Kapteyn Astronomical Institute, University of Groningen, PO Box 800, NL-9700 AV Groningen, the Netherlands\\
$^{2}$School of Physics and Astronomy, University of Southampton, Southampton, SO17 1BJ, UK\\
$^{3}$Instituto Argentino de Radioastronom\'ia (CCT La Plata, CONICET; CICPBA; UNLP), C.C.5, (1894) Villa Elisa, Buenos Aires, Argentina\\
$^{4}$Key Laboratory of Particle Astrophysics, Institute of High Energy Physics, Chinese Academy of Sciences, 19B Yuquan Road, Beijing 10049,\\ People's Republic of China
}

\date{Accepted 2022 May 23. Received 2022 May 20; in original form 2021 December 6}

\pubyear{2015}

\begin{document}
\label{firstpage}
\pagerange{\pageref{firstpage}--\pageref{lastpage}}
\maketitle

\begin{abstract}
We study the properties of the type-C quasi-periodic oscillation (type-C QPO) of MAXI~J1348--630 during its 2019 outburst and reflare with \textit{NICER}. This is the first time that the evolution of the properties of type-C QPOs are studied during an outburst reflare. We found that the properties of the type-C QPO during the reflare are similar to those of type-C QPOs observed in other black-hole systems during outburst. This suggests that the physical processes responsible for type-C QPOs are the same in a reflare and in an outburst. We also found that the FWHM of a high-frequency broadband component observed during the reflare changes significantly with energy. We studied the energy-dependent fractional rms amplitude and phase lags of the type-C QPO from 0.5 keV to 12 keV. We found that the fractional rms amplitude increases up to 2--3 keV and then remains approximately constant above this energy, and the lag spectra of the type-C QPO are hard. We discuss the dependence of the fractional rms amplitude and phase lags with energy in the context of Comptonisation as the radiative mechanism driving the QPO rms and lag spectra.
\end{abstract}




\section{Introduction}

The spectral and timing properties of black-hole low-mass X-ray binaries (BH LMXBs) vary in a continuous manner during outburst \citep[e.g., ][]{Remillard06}. Depending on these properties, two main spectral states are defined \citep[see e.g., ][]{Tanaka89, vanderKlis94}: the low/hard state (LHS) and the high/soft state (HSS). In the LHS, the energy spectrum is dominated by a Comptonised component (hereafter called corona) described by a hard power-law. In the HSS, on the other hand, the energy spectrum is dominated by a thermal component described by a multi-colour disc blackbody. There are two intermediate states between the LHS and HSS: the \textit{hard-intermediate state} (HIMS) and the \textit{soft-intermediate state} (SIMS) with properties in between those of the LHS and the HSS \citep[][]{Homan05}. During a full outburst, BH LMXBs evolve through all these states in the order: $LHS\rightarrow HIMS \rightarrow SIMS \rightarrow HSS \rightarrow SIMS \rightarrow HIMS \rightarrow LHS$. Sometimes, some BH LMXBs undergo \textit{Failed-Transition} outbursts (hereafter FT outbursts), in which these systems do not reach the HSS and either remain in the LHS during the whole outburst \citep[e.g., ][]{Hynes00b, Brocksopp01, Belloni02,Brocksopp04,Curran13} or evolve to the HIMS to return immediately to the LHS \citep[e.g., ][]{Zand02b,Capitanio09a,Ferrigno12}.

The timing properties of BH LMXBs also change with the evolution of the source along the different spectral states. In the LHS the power-density spectrum (PDS) is characterised by a strong broadband noise component with a fractional rms amplitude of 30\%--50\% \citep[e.g.][]{Mendez97,Belloni05,Remillard06, Munoz-Darias11, Motta16} while in the HSS the broadband fractional rms of BHs is generally less than 5\% \citep{Mendez97}. The fractional rms amplitude in the HIMS and the SIMS ranges between 5\% and 20\% \citep[e.g., ][]{Belloni10, Munoz-Darias11, Motta12}. 

In addition to the broadband noise component, quasi-periodic oscillations (QPOs) are detected in the PDS of a BH LMXB during an outburst \citep[e.g., ][]{Psaltis99, Nowak00, Casella04, Belloni05, Belloni10}. Based on the frequency range, two groups of QPOs are identified in BH LMXBs: low-frequency QPOs (LF QPOs), with frequencies ranging from a few mHz to 30 Hz \citep[e.g., ][]{Belloni02a, Casella04, Remillard06c}, and high-frequency QPOs (HF QPOs), with frequencies up to $\sim$500 Hz \citep[e.g., ][]{Morgan97, Remillard99a, Belloni01, Homan01, Remillard02,  Altamirano12}. LF~QPOs can be classified into three main groups: type-A, -B and -C based on the combined properties of the QPO and the broadband noise components \citep[e.g., ][]{Wijnands99, Casella04, Casella05}. 
Type-C QPOs are detected in the LHS and the HIMS and are characterised by a strong and narrow peak with a centroid frequency ranging from 0.01 Hz to 30 Hz superposed to a strong, 15--30\% fractional rms amplitude, broadband noise component \citep[e.g., ][]{Casella04, Belloni05}. A subharmonic and a second harmonic peak are usually also present in the PDS \citep[][]{Casella05}. Type-B QPOs are detected in the SIMS and they show centroid frequencies in the 1--7 Hz frequency range \citep[e.g, ][]{Wijnands99, Casella04, Casella05} on top of a weaker broadband noise than that seen for type-C QPOs. Type-A QPOs are also detected in the SIMS, have centroid frequencies in the 6.5--8 Hz frequency range and are broader than type-B and type-C QPOs \citep{Wijnands99,Casella04,Belloni14}.

In this work we focused on the analysis of type-C QPOs, which are the most common of the three types of QPO. The physical mechanism driving the type-C QPOs is not clear yet. Dynamical models proposed to explain the frequency of the type-C QPOs are based on two mechanisms: instabilities of the accretion flow \citep[e.g., ][]{Tagger99, Titarchuk99, Titarchuk04, Cabanac10} and geometrical effects, as the Lens-Thirring (L-T) precession \citep[e.g., ][]{Stella98, Stella99, Schnittman06, Ingram09, Ingram11}, the latter being one of the most promising models. In the L-T model, the type-C QPO is produced by the L-T precession of a radially extended region of the inner hot flow in truncated accretion disc models. Recent results have reinforced the geometric interpretation for the type-C QPO. \citet{Motta15} and \citet{Heil15} found evidence that the rms amplitude of the QPO depends on the orbital inclination, while \citet{Ingram16} and \citet{Ingram17} found that the centroid energy and the reflection fraction of the iron line, respectively, are modulated at the QPO phase.

The study of the energy-dependent timing properties of QPOs, such as the fractional rms amplitude and the lags between different energy bands, can help us to understand the radiative process behind QPOs. The study of the rms spectra of type-C QPOs has been done for 20 years \citep[e.g., ][]{Tomsick01c, Casella04, Casella05, Rodriguez04, Rodriguez04a, Sobolewska06, Axelsson16, Zhang17b, Zhang20b, Karpouzas21, Ma21}. In general, the fractional rms amplitude of type-C QPOs increases with energy up to 10--20 keV and remains more or less constant above that up to $\sim$30--50 keV \citep[e.g., ][]{Casella04, Zhang17b, Zhang20b}. In the case of the type-C QPO in MAXI~J1820+070, \citet{Ma21} found that the fractional rms amplitude remains more or less constant at $\sim$10\% up to 200 keV. On the other hand, \citet{Tomsick01c} found that, in some cases, after reaching a maximum at $\sim$20 keV, the rms of the QPO decreases slightly at higher energies.

The high fractional rms values of type-C QPOs at energies above $\sim$20--30 keV suggests that the QPO emission is dominated by the Comptonised component and that the radiative mechanism of type-C QPOs cannot be related to the disc, which supports models that identify the type-C QPOs as oscillations in the physical properties of the corona \citep[e.g., ][]{Lee01,Kumar14, Karpouzas20}. It has been recently shown that these models can reproduce the rms and lag spectra of BH systems \citep[][]{Garcia21, Karpouzas21, Zhang22, Garcia22} and NS systems \citep[][]{Karpouzas20}.

The study of the energy-dependent lags of QPOs, as for the rms spectra, started around 20 years ago \citep[][]{Vaughan94, Vaughan97}. Since then, this technique has become a very powerful tool to study the radiative properties of the X-ray variability. Lags represent the time delay between photons of two different energy bands. If the high-energy photons are delayed with respect to those from low-energy bands, the lag is defined as positive and are called hard lags. If, on the contrary, the low-energy photons are delayed with respect to those from high-energy bands, the lag is defined as negative and are called soft. Both soft and hard lags have been observed for type-C QPOs of BH LMXBs \citep[e.g., ][]{Reig00, Cui00a, Casella04, Munoz-Darias10, Pahari13, Zhang17b, Jithesh19, Zhang20b}. \citet{vandenEijnden17} found evidence in a sample of 15 BH LMXBs that the phase lags at the frequency of type-C QPOs depend on the source inclination. High-inclination sources show soft lags at high QPO frequencies, while low-inclination sources display hard lags. All sources display hard lags at low QPO frequencies. However, GRS~1915+105 does not present this behaviour \citep[][]{Zhang20b}


Some models have been proposed to explain the phase lags of QPOs \citep[e.g., ][]{Lee98, Kotov01, Ingram09, Shaposhnikov12, Misra13, Ingram16}. A Comptonisation model to explain the hard lags of the broadband component of Cyg~X--1 was presented by \citet{Miyamoto88}. Later, \citet{Nobili00} presented a Comptonisation model that explains both the hard and soft lags of the QPO of GRS~1915+105 in terms of variations of the truncation radius of the accretion disc. \citet{Kumar14} and \citet{Karpouzas20} proposed a Comptonisation model taking into account feedback between the corona and the accretion disc \citep[][]{Lee01}, in which a fraction of the up-scattered photons in the corona impinges back on to the accretion disc producing a time delay between the hard photons from the corona and those from the accretion disc. This generates soft lags that are observed in some systems \citep[e.g., ][]{Reig00, Pahari13, vandenEijnden17, Zhang20b}. 

During the decay of an outburst, or before reaching the end of it, sometimes the source rebrightens reaching X-ray luminosities one or two orders of magnitude lower than those at the peak of the outburst. These phenomena are known in the literature as \textit{``reflares''}, \textit{``rebrightenings''}, \textit{``rebursts''}, \textit{``echo-outbursts''} or \textit{``mini-outbursts''}, and have been observed  in a few sources \citep[e.g., ][]{Callanan95, Altamirano11a, Jonker12, Patruno16, Cuneo20, Zhang20, Zhang20a}.

The spectral and timing properties of low-mass X-ray binaries during a reflare have been studied in some BH LMXBs \citep[e.g., ][]{Kuulkers96, Kuulkers98, Tomsick03, Homan13, Yan17, Cuneo20, Zhang20, Stiele20} and neutron-star LMXBs \citep[][]{Simon10, Patruno16, Bult19}. \citet{Yan17} studied the energy spectra of the BH LMXB GRS~1739--278 during the reflares of its 2014 outburst and found that the source underwent spectral transitions during the reflares and showed hysteresis loops in the hardness-intensity diagram (HID). In addition, these authors found that the peak luminosity of the HSS and the luminosity at which the hard-to-soft state transition occurred follow a correlation previously observed for main outbursts. \citet{Cuneo20} studied the spectral and timing properties of the BH~LMXB MAXI~J1535--571 during its reflares and also found spectral transitions and loops as those found in GRS~1739--278. Alternatively, the BH systems MAXI~J1659--152 \citep[][]{Homan13}, XTE~J1650--500 \citep[][]{Tomsick04}, MAXI~J1348--630 \citep[][]{Zhang20} and MAXI~J1820$+$070 \citep[][]{Stiele20} showed reflares during which these four sources remained in the LHS during the whole reflare. This behaviour is similar to that seen in FT outbursts \citep[e.g., ][]{Hynes00b, Brocksopp01, Belloni02,Brocksopp04,Curran13}. Moreover, the BH~LMXBs MAXI~J1348--630 and MAXI~J1820$+$070 showed type-C QPOs during their reflares \citep[][ respectively]{Zhang20, Stiele20}.

The fact that they are observed in systems with different types of compact objects suggests that reflares are related to the properties of the accretion disc and are independent of the nature of the compact object. In addition, the similarities between the spectral and timing properties between the outbursts and reflares points out to the same physical origin for both phenomena \citep[e.g., ][]{Patruno16, Cuneo20}. If the trigger of reflares is an instability in the accretion disc, this is problematic for the the disc instability model \citep[DIM; see ][ for a review]{Lasota01}. This model needs a large amount of matter at the outer accretion disc to trigger an instability and, at the same time, predicts that the disc is depleted from matter at the end of the outburst, when the reflares are produced. Although several models have been proposed to explain the reflares \citep[e.g., ][]{Chen93, Kuulkers94, Hameury00, Hameury00a, Zhang19}, none of them have been proven yet. Therefore it is very important to study the properties of the reflares, in order to understand their physical mechanism.

MAXI~J1348--630 is an X-ray binary discovered with MAXI on January 2019 \citep[][]{Yatabe19, Tominaga20}. The source was also detected with \textit{Swift} \citep[][]{D'Elia19a, D'Elia19b}, \textit{NICER} \citep[][]{Sanna19}, \textit{ATCA} \citep[][]{Russell19} and \textit{INTEGRAL} \citep[][]{Cangemini19a}. Based on its spectral and timing properties studied with \textit{NICER} data, \citet{Sanna19} suggested that the compact object in this system is a BH. Later on, a more detailed study of the evolution of these properties during the whole outburst using \textit{NICER}
allowed to reinforce the identification of MAXI~J1348--630 as a BH candidate \citep[][]{Zhang20}.

In this paper, we present the study of the properties of the type-C QPOs of MAXI~J1348--630, focusing on the energy-dependent fractional rms amplitude and phase lags of the type-C QPO. This is the first study showing the evolution of the timing properties of type-C QPOs during a reflare of a BH LMXB. In section 2 we describe the observations and data analysis. In section 3.1 we show the properties of the PDS of the observations of MAXI~J1348--630 showing a type-C QPO and the properties of this QPO. In section 3.2 and 3.3, respectively, we study the fractional rms and phase lag spectra of the type-C QPO. Finally, in section 4 we compare the properties of type-C QPO in the reflare with those of type-C QPOs in other systems and we discuss the energy-dependent fractional rms amplitude and phase lags in terms of Comptonisation mechanisms.

\section{Observations and data analysis}

MAXI J1348--630 was observed 253 times since 26 January of 2019 with \textit{NICER} \citep[][]{Gendreau12} on an almost daily basis (ObsID $1200530101-3200530232$). We analysed the data using the software HEASOFT version 6.26 and NICERDAS version 6.0. The CALDB version used in this project was 20190516. We applied standard filtering and cleaning criteria. We included the data when the dark Earth limb angle was $>15\degr$, the pointing offset was $<54\arcsec$, the bright Earth limb angle was $>30\degr$, and the International Space Station (ISS) was outside the South Atlantic Anomaly (SAA). Although 52 detectors of \textit{NICER} were active during all the observations of MAXI J1348--630, we removed data from detectors 14 and 34, since they occasionally show episodes of increased electronic noise. We also omitted detectors 10--17 from MJD 58672 (ObsID 2200530169) to MJD 58687 (ObsID 2200530181) due to a temporary instrument anomaly. From MJD 58509 (ObsID 1200530101) to MJD 58524.8 (ObsID 1200530109) MAXI J1348--630 was observed with a pointing offset of 2.2 arcmin (${\rm RA}=13^{\rmn h} 47^{\rmn m} 55^{\rmn s}$ and ${\rm DEC}=-63\degr 15\arcmin 34\arcsec$).

To create the long-term light curve and hardness-intensity diagram \citep[HID, ][]{Homan01,Remillard06,Belloni06} we first extracted a background subtracted energy spectra for each ObsID using the nibackgen3c50 tool. We then obtained the count rate of the source in the 0.5--12 keV, 2--3.5 keV and 6--12 keV energy bands for each ObsID. We defined the intensity as the background-subtracted count rate in the 0.5--12 keV energy range and the hardness ratio as the ratio between the background-subtracted count rate of the 6--12 keV and 2--3.5 keV energy bands (one point per ObsID).

For the Fourier timing analysis we constructed Leahy-normalised \citep[][]{Leahy83} power spectra using data segments of 32.768 seconds and a time resolution of 250$\mu$s. The minimum frequency was $\sim$0.03 Hz and the Nyquist frequency was 2000 Hz. We then averaged the power spectra per ObsID, subtracted the Poisson noise based on the average power in the $500-2000$ Hz frequency range, and converted the power spectra to fractional rms \citep[][]{vanderKlis95a}. To fit the power spectra we used a multi-Lorentzian function. We used the characteristic frequency of the Lorentzians defined in \citet{Belloni02a}, $\nu_{max}=\sqrt{\nu_{0} + (FWHM/2)^2} = \nu_{0} \sqrt{1+1/4Q^{2}}$, where $FWHM$ is the full width at half maximum, $\nu_{0}$ the centroid frequency of the Lorentzian and the quality factor $Q$ is defined as $Q = \nu_{0}/FWHM$.

In order to identify the QPO, we inspected the PDS of each observation. If there was only one narrow component ($Q > 0$), we identified it as the QPO. If there were two or more narrow Lorentzians, we assumed that the fundamental QPO was the strongest and narrowest Lorentzian and if the other narrow Lorentzians were at frequencies $\sim$1/2 or $\sim$2 times the frequency of the QPO, we identified them as the subharmonic or the harmonic, respectively. If the other narrow Lorentzians were not at frequencies corresponding to harmonics or subharmonics, we identified them as narrow peaks. In order to reinforce the identification of the QPO fundamental, we studied the relation between the frequency of the QPO and the 0.01--64 Hz broadband fractional rms amplitude in the PDS \citep[][]{Casella04, Motta11}. This relation allows us to check whether the QPO is the same component in all the observations and to identify the type of QPO according to the broadband fractional rms amplitude. In order to identify the broad components in the PDS, we compared them with those observed in other systems \citep[e.g., ][]{Belloni02a, Altamirano05, Altamirano08, Klein-Wolt08}.

To obtain the rms spectrum, we carried out the procedure described above in the following energy bands: $0.5-1.0$ keV, $1.0-1.5$ keV, $1.5-2.0$ keV, $2.0-3.0$ keV, $3.0-4.0$ keV, $4.0-6.0$ keV, $6.0-8.0$ keV and $8.0-12.0$ keV.

We present very briefly a new method to obtain the individual phase lag spectrum of a component of the PDS; the full explanation of the method will be presented in a forthcoming paper (Peirano et al. in prep). In general, the method used to estimate the lag of the QPO is to average the frequency-dependent Real and Imaginary parts of the cross spectrum within the FWHM of the QPO \citep[e.g., ][]{Vaughan97, Nowak99, Uttley14}. This is in principle correct when the QPO is strong enough to dominate the PDS in the frequency range of interest. However, it can happen that the QPO is not as strong, and thus the phase lags in the frequency range of interest can be due to multiple PDS components, and thus the averaged lags do not represent the lags of the QPO itself. Assuming that the QPO is independent of the broadband variability (and additive), the method we used here allows us to overcome this issue.

In order to estimate the lag of the Lorentzian associated to the QPO, we first produce the cross spectra using the selected subject bands listed above for the rms. Then, using the same multi-Lorentzian model used to fit the associated PDS, we simultaneously fit the Real and Imaginary parts of each cross spectrum in \textsc{XSPEC}, fixing the centroid frequencies and FWHM of each Lorentzian in the model, but letting their normalisations free to vary. We then can calculate the lag of the fundamental QPO Lorentzian (and of every other individual component) by taking the arctan of the ratio of the integrals of the Lorentzians associated to the Imaginary and Real parts of the QPO, respectively, $\phi = atan(Im(QPO)/Re(QPO))$. In reality, we parametrise the normalisation of the Lorentzian in the Imaginary part of the cross spectrum as the normalisation of the Real part $\times tan(\phi)$, and let $\phi$ free. In order to estimate the errors, we obtained the 1$\sigma$ errors using XSPEC error command.

\section{Results}

\subsection{PDS of the observations with type-C QPOs}

\begin{figure*}
    \centering
    \includegraphics[width=\textwidth]{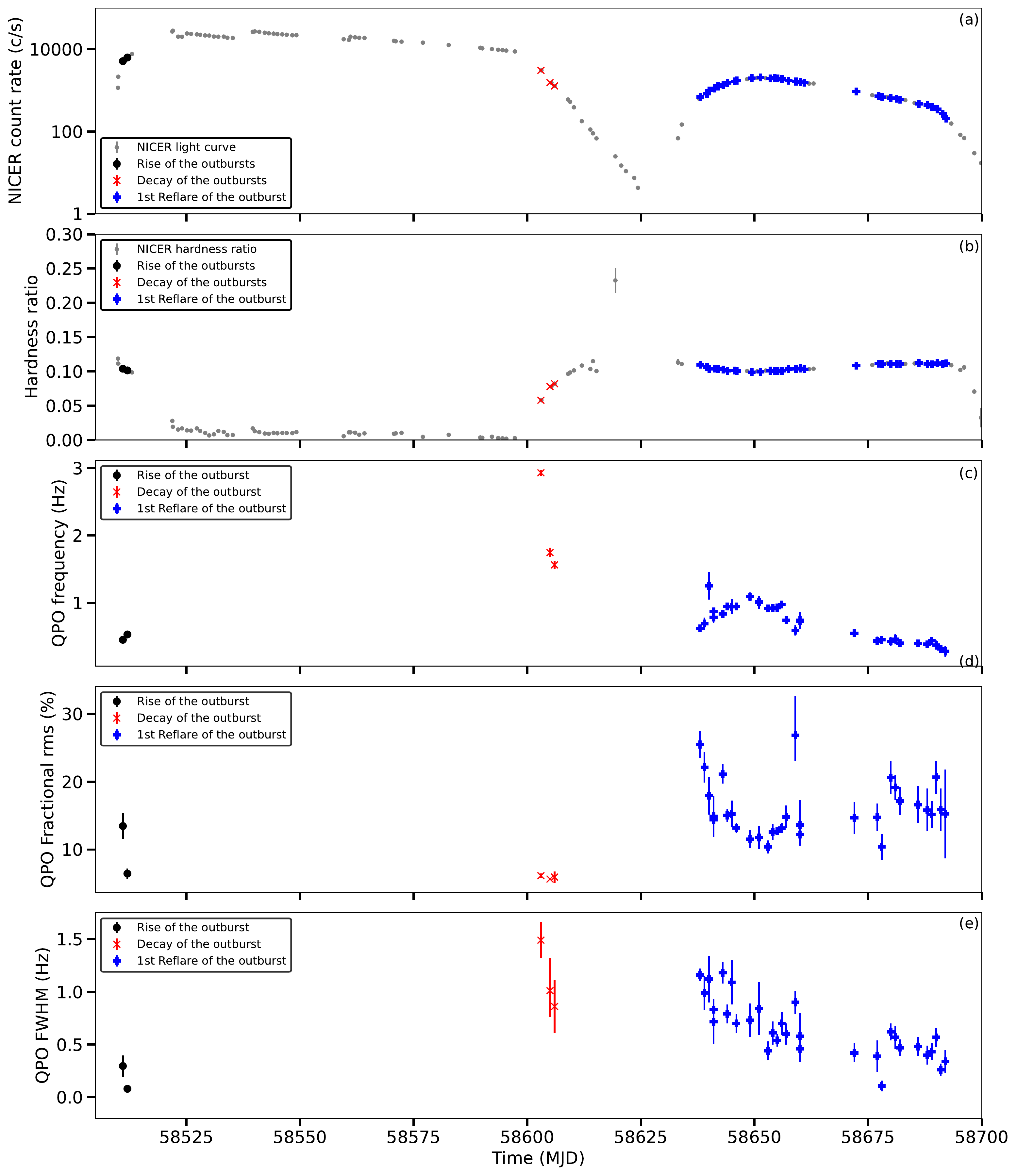}
    \caption{Panel (a):  \textit{NICER} light curve of MAXI~J1348--630. Panel (b): Temporal evolution of the hardness ratio of MAXI~J1348-630. Panel (c): Temporal evolution of the QPO frequency. Panel (d): Temporal evolution of the fractional rms amplitude of the QPO. Panel (e): Temporal evolution of the FWHM of the QPO. Black, red and blue symbols represent the data corresponding to the rise, the decay and the first reflare of the outburst, respectively. Grey symbols represent the full \textit{NICER} light curve (panel (a)) and the full temporal evolution of the hardness ratio (panel (b)).}
    \label{fig:lc_hardness}
\end{figure*}

\begin{figure}
    \centering
    \includegraphics[width=\columnwidth]{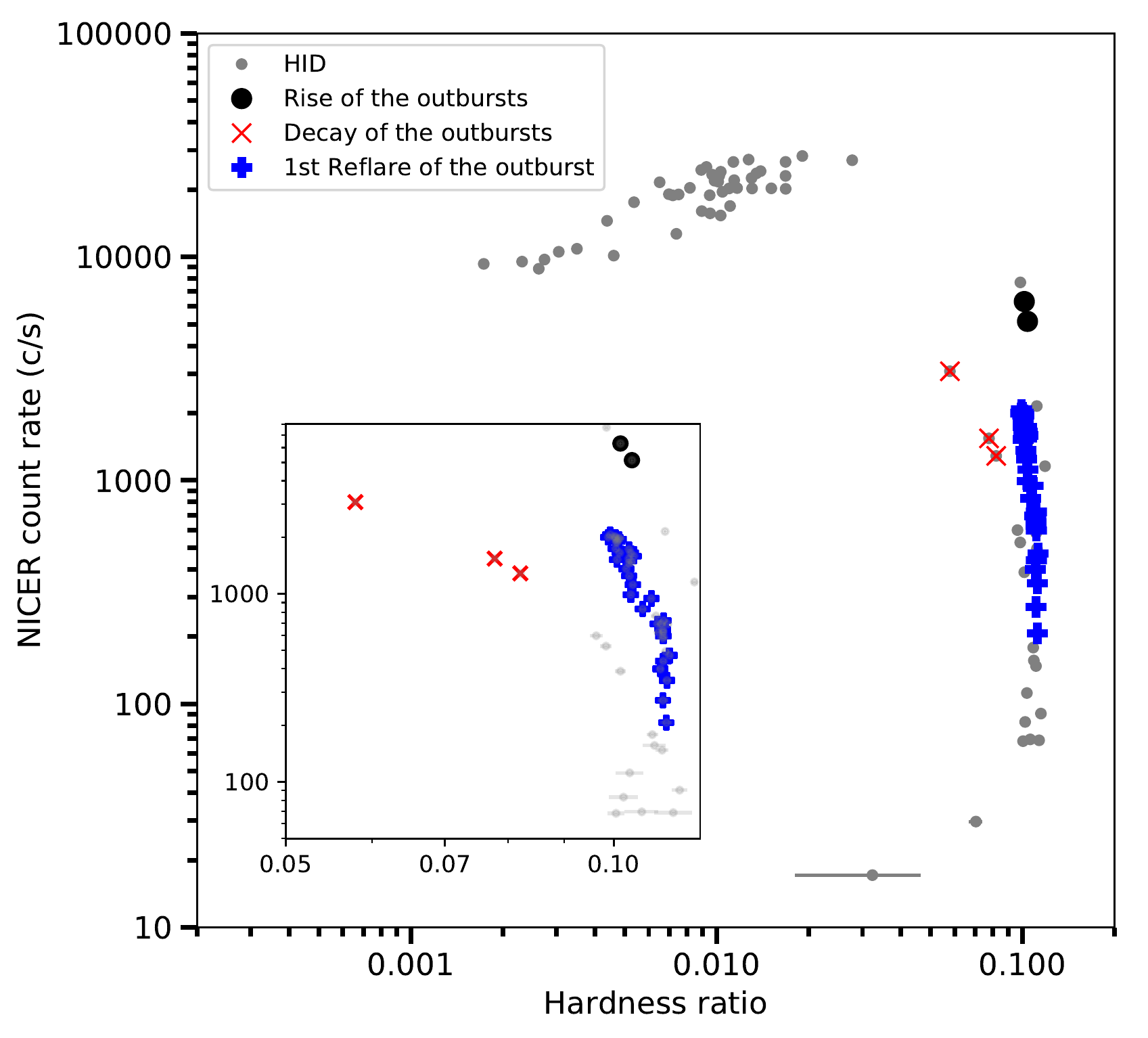}
    \caption{HID of MAXI~J1348--630. Grey: Data points for the full outburst and reflares. Colours and symbols are the same as in Fig. \ref{fig:lc_hardness}.}
    \label{fig:hid}
\end{figure}

The evolution of the PDS of MAXI~J1348--630 along the main outburst and the first reflare is described on Section 3.3 of \citet{Zhang20}. We focused on the study of the observations in which a type-C QPO was observed. Table \ref{Tab:nicer_obsid} contains the list of observations with type-C QPOs and their location in the light curve and hardness ratio are shown in panels (a) and (b) of Fig. \ref{fig:lc_hardness}, respectively. Fig. \ref{fig:hid} shows the HID of the system.

\begin{figure*}
    \centering
    \includegraphics[width=\textwidth]{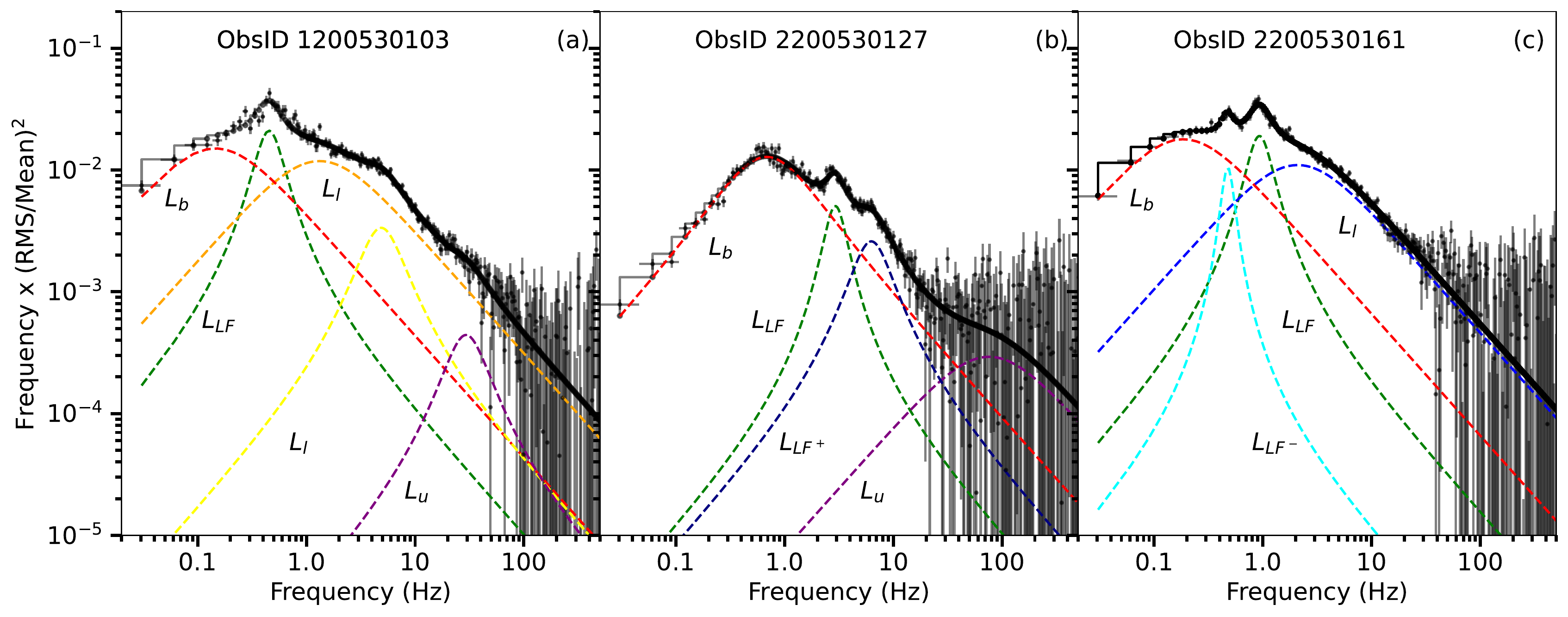}
    \caption{Three representative power spectra of observations with type--C QPOs of the outburst of MAXI~J1348--630 during the rise (panel (a)), decay (panel (b)) and first reflare of the outburst (panel (c)). Dashed lines represent the best fit Lorentzians. The different components are identified with a name based on previous works on time variability in low-mass X-ray binaries.}
    \label{fig:power}
\end{figure*}

\begin{figure}
    \centering
    \includegraphics[width=\columnwidth]{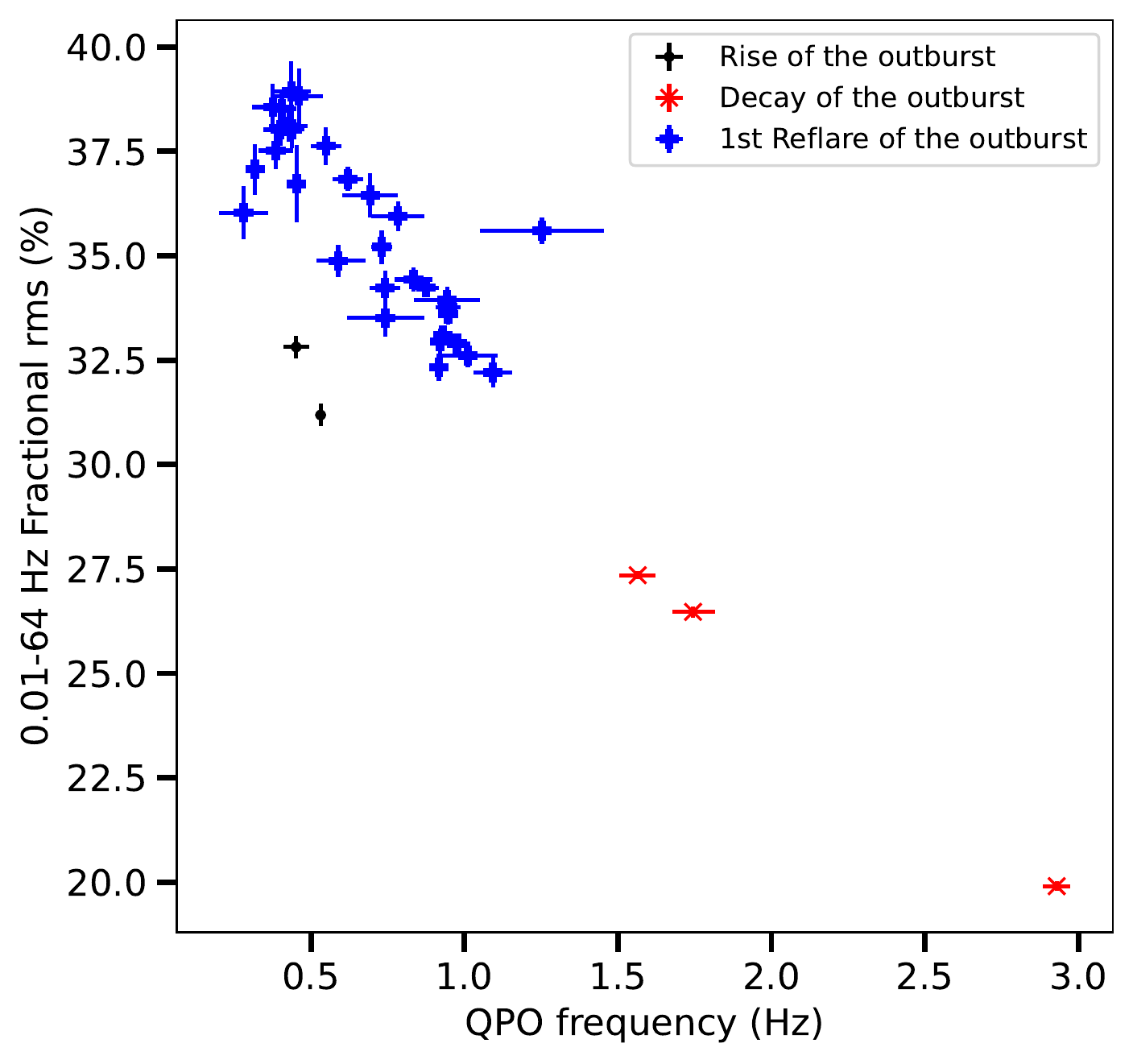}
    \caption{Fractional rms amplitude in the 0.01--64 Hz frequency range vs QPO frequency. Colours and symbols are the same as in Fig. \ref{fig:lc_hardness}.}
    \label{fig:rms_vs_freq}
\end{figure}

Fig. \ref{fig:power} shows three representative PDS of observations showing a type-C QPO. During the rise of the main outburst, a type-C QPO is detected significantly in two observations: ObsIDs 1200530103 and 1200530104 (Panel (a) of Fig. \ref{fig:power}). The PDS of these two observations were fitted with two zero-centred Lorentzian and three narrow Lorentzian. During the decay of the main outburst, a type-C QPO is found in three observations: ObsIDs 220053127--220053129 (panel (b) of Fig. \ref{fig:power}). These observations were fitted with one zero-centred and three narrow Lorentzians. Finally, during the first reflare of MAXI~J1348--630, we found a type-C QPO in several observations (see Table \ref{Tab:nicer_obsid} for the list of ObsIDs and panel (c) in Fig. \ref{fig:power} for a representative PDS). The PDS of these observations were well fitted with two zero-centred and one narrow Lorentzian. A second narrow Lorentzian is significantly detected at lower frequencies than the previous narrow component in some observations during the reflare. 

\begin{figure*}
    \centering
    \includegraphics[width=0.9\textwidth]{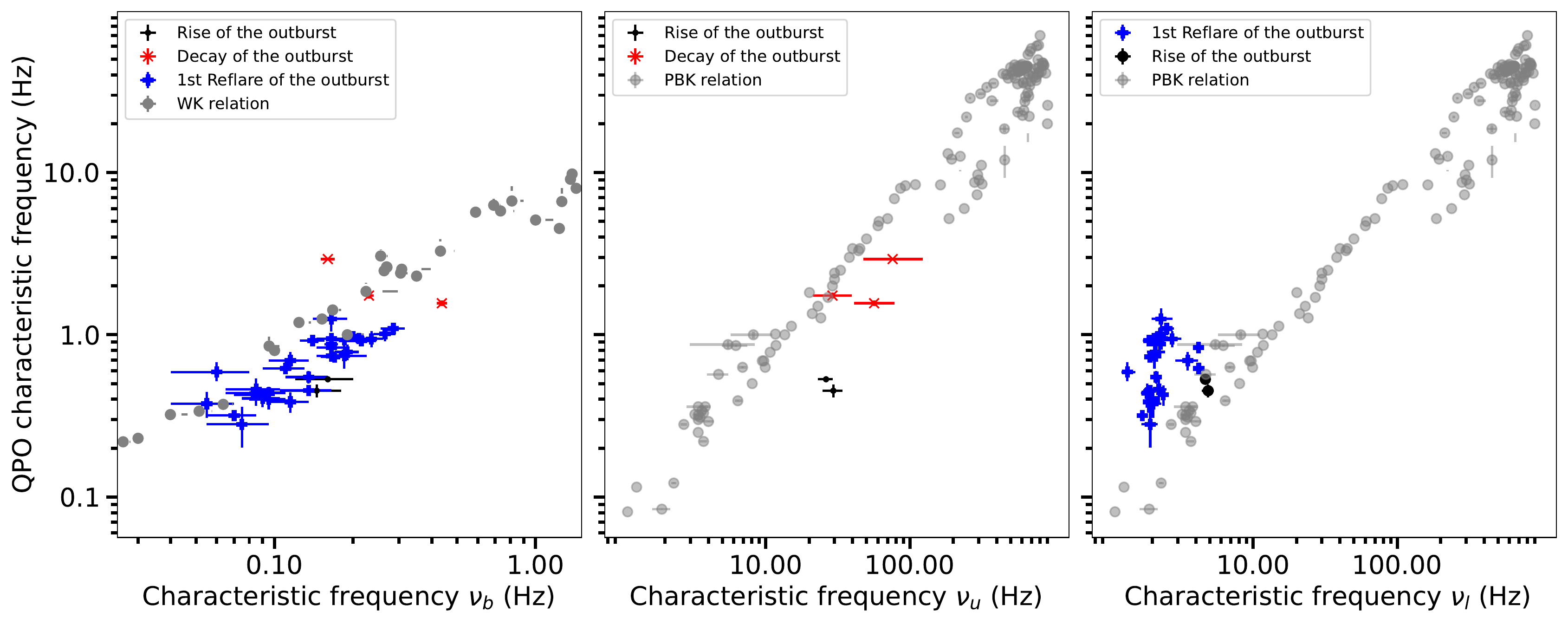}
    \caption{Relation of the frequencies of $L_{LF}$ with $L_b$ (left panel); $L_u$ (middle panel) and $L_l$ (right panel). Black, red and blue symbols are the same as in Fig. \ref{fig:lc_hardness}. Grey symbols represent the PBK relation with data from \citet{Psaltis99}, \citet{Nowak00}, \citet{Homan01}, \citet{Belloni02a} and \citet{Nowak02} in the middle panel. Grey symbols represent the WK relation as shown in \citet{Wijnands99a} in the left and right panels.}
    \label{fig:freq_corr}
\end{figure*}

We identified the type-C QPO as the narrowest components in all the PDS (see Table \ref{Tab:nicer_obsid} for their frequencies). Panels (a), (b) and (c) of Fig. \ref{fig:power} show three representative QPOs peaking at $\sim$0.45 Hz, $\sim$2.93 Hz and $\sim$0.93 Hz, respectively. In order to reinforce our identification of the QPO, we plot the 0.01--64 Hz fractional rms amplitude vs. the frequency of the QPO in Fig. \ref{fig:rms_vs_freq}. We found that both quantities are anti-correlated. This relation is similar to that found by \citet{Casella04} for the type-C QPOs of XTE~J1859$+$226. The smooth shape of the relation confirms that our identification of the QPO is correct. In addition, we see that the broadband fractional rms of the observations with QPOs is always higher than $\sim$20\%, supporting our identification of the QPOs as type-C. Following \citet{Belloni02a} and \citet{Klein-Wolt08} we called this component $L_{LF}$ in the three panels, where $LF$ means \textit{low-frequency}. The frequency of this component is called $\nu_{LF}$. During the decay of the outburst we found a component peaking at frequencies 2 times the frequency of $L_{LF}$ (see the light blue component in panel (b) of Fig. \ref{fig:power}). We identified this component as the second harmonic of the QPO. In some observations of the reflare we also found another narrow component peaking at half of the QPO frequency (see the light blue component in panel (c) of Fig. \ref{fig:power}). We identified this component as the sub-harmonic of the QPO. Following the aforementioned nomenclature, we called the harmonic and sub-harmonic $L_{LF}^{+}$ and $L_{LF}^{-}$, respectively. Their frequencies are called $\nu_{LF}^{+}$ and $\nu_{LF}^{-}$, respectively.
Regarding the broadband components and the other narrrow components, we identified them by comparing them with those shown in previous studies \citep[e.g., ][]{Belloni02a, Altamirano05, Altamirano08, Klein-Wolt08}. Broadband components can also be seen in Fig. \ref{fig:power}. During the rise of the outburst, the fits required two broad components at low frequencies: $L_b$ and $L_{h}$. We identified the component corresponding to the break frequency as the one peaking at $\sim$0.15 Hz in panel (a) of Fig. \ref{fig:power}. We then called it $L_b$ and its frequency $\nu_{b}$. We identified this component by studying its correlation with the frequency of $L_{LF}$ (see the left panel of Fig. \ref{fig:freq_corr}). The frequency of $L_b$ is correlated with $\nu_{LF}$ and follows the WK relation \citep[][]{Wijnands99a}, reinforcing our identification of $L_b$. We called the component peaking at $\sim$0.5 Hz $L_{h}$ and its frequency $\nu_{h}$. In addition to $L_{LF}$ and $L_b$, we also identified the components $L_u$ and $L_l$ in the PDS studied in this project. As we show below (see Fig. \ref{fig:freq_corr}), these components are consistent with $L_l$ and $L_u$ in the PBK relation \citep[][]{Psaltis99}. Examples of these components can be seen in Fig. \ref{fig:power}.

We also studied the relation between the frequencies of the components $L_{b}$, $L_{u}$ and $L_{l}$ with the frequency of $L_{LF}$. Fig. \ref{fig:freq_corr} shows these relations. We found that $\nu_{LF}$ is correlated with $\nu_{b}$ (left panel), $\nu_{u}$ (middle panel) and $\nu_l$ (right panel). The relation of $\nu_{LF}$ with $\nu_b$ and $\nu_u$ lie slightly below the WK \citep[][]{Wijnands99a} and PBK \citep[][]{Psaltis99} relations, respectively. Alternatively,, the relation of $\nu_{LF}$ with $\nu_l$ lies slightly above the PBK during the reflare.  

\subsection{Evolution of the QPO parameters with time, hardness ratio and \textit{NICER} count rate}

The parameters of the QPO evolved with time, hardness ratio and the X-ray count rate. Panels (c), (d) and (e) of Fig. \ref{fig:lc_hardness} show the temporal evolution of the $\nu_{LF}$, the fractional rms amplitude and the FWHM of the type-C QPO, respectively. The frequency, the fractional rms amplitude and the FWHM of the QPO range from $\sim$0.29 Hz to $\sim$2.92 Hz, $\sim$6\% to $\sim$ 26\% and $\sim$0.08 Hz to $\sim$1.49 Hz, respectively. During the rise of the outburst (black symbols in panel (c) of Fig. \ref{fig:lc_hardness}), $\nu_{LF}$, the fractional rms and the FWHM decrease with time. However, we need to be cautious about this result, since we only have two measurements during the rise of the outburst. During the decay of the outburst (red symbols in Fig. \ref{fig:lc_hardness}), $\nu_{LF}$ decreases from $\sim$2.9 to $\sim$1.6 Hz, the fractional rms amplitude remains constant in the three observations at $\sim$6\%, and the FWHM decreases from $\sim$1.5 Hz to $\sim$0.9 Hz. During the reflare (blue symbols in Fig. \ref{fig:lc_hardness}), $\nu_{LF}$ increases from $\sim$0.6 Hz (MJD 58638) to $\sim$1.10 Hz (MJD 58649) and then decreases to $\sim$0.30 Hz (MJD 58692), corresponding to the last observation of the reflare showing a type-C QPO. The fractional rms amplitude decreases from $\sim$25\% (MJD 58638) to $\sim$12\% at the peak of the reflare and then remains in the range 11--26\% at the decay of the reflare. The FWHM of the QPO decreases along the reflare.

\begin{figure}
    \centering
    \includegraphics[width=\columnwidth]{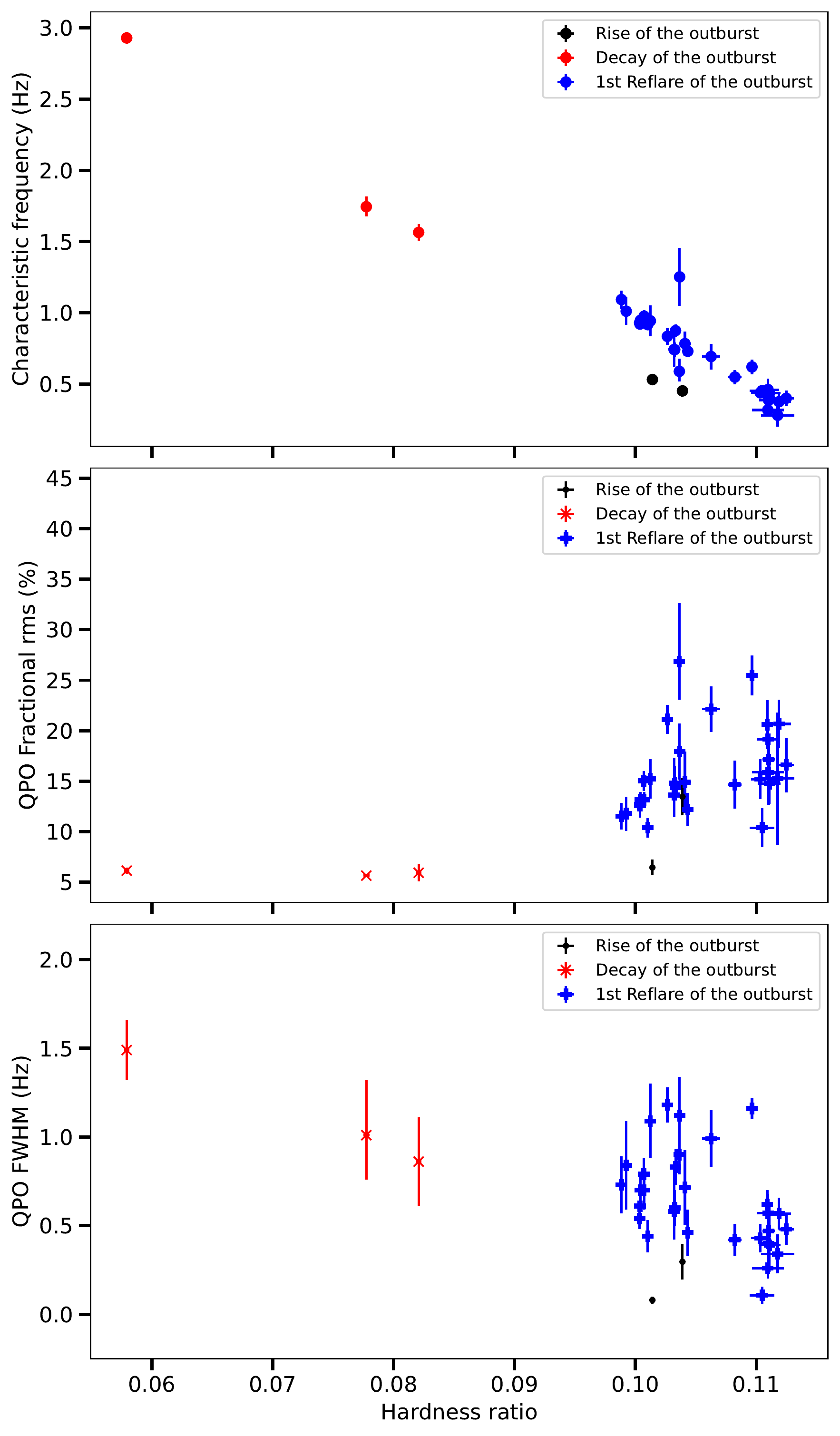}
    \caption{Frequency (top panel), fractional rms amplitude (middle panel) and FWHM (bottom panel) of the type-C QPO vs. the hardness ratio.}
    \label{fig:param_vs_hardness}
\end{figure}

Fig. \ref{fig:param_vs_hardness} shows the evolution of the frequency (upper panel), the fractional rms amplitude (middle panel) and FWHM (lower panel) of $L_{LF}$ with hardness ratio. Both $\nu_{LF}$ and the FWHM are anti-correlated with hardness ratio during the whole outburst and reflare. However, the relation between the FWHM and the hardness ratio is less clear during the reflare due to some scatter. The evolution of the fractional rms amplitude is different in each phase of the outburst. During the rise of the outburst (black symbols in the middle panel of Fig. \ref{fig:param_vs_hardness}), although there are only two measurements in the rise of the outburst, the fractional rms amplitude appears to increase with hardness ratio within errors. During the decay of the outburst (red symbols in the middle panel of Fig. \ref{fig:param_vs_hardness}), the fractional rms amplitude remains approximately constant with hardness ratio. The fractional rms amplitude appears to increase slightly with the hardness ratio during the reflare (blue symbols in the middle panel of Fig. \ref{fig:param_vs_hardness}).

\begin{figure}
    \centering
    \includegraphics[width=\columnwidth]{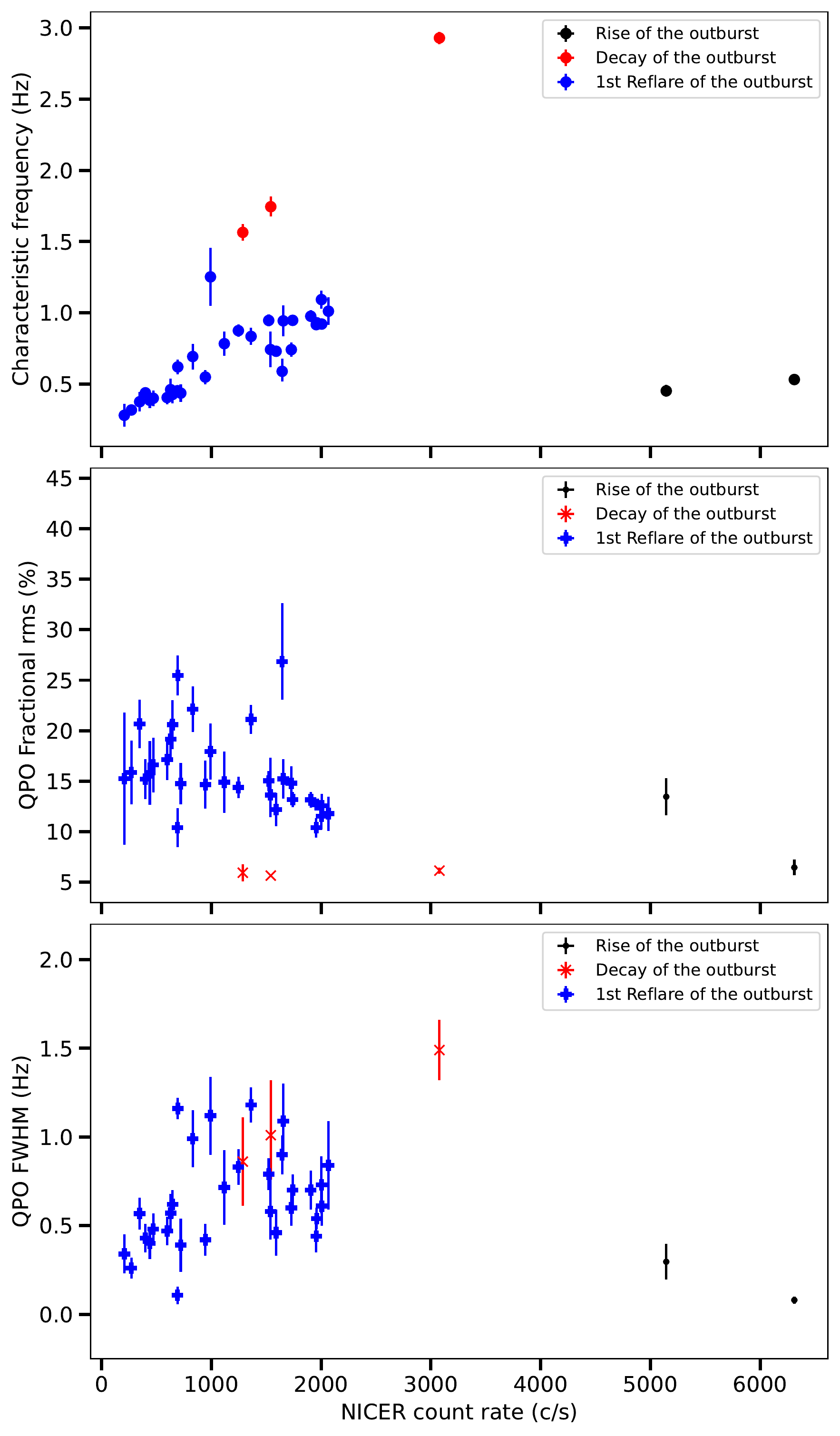}
    \caption{Frequency (top panel), fractional rms amplitude (middle panel) and FWHM (bottom panel) of the type-C QPO vs. the count rate.}
    \label{fig:param_vs_rate}
\end{figure}

Fig. \ref{fig:param_vs_rate} shows the relation of the frequency (top panel), fractional rms (middle panel) and FWHM (lower panel) of the QPO with count rate. The frequency appears to remain approximately constant with count rate within errors during the rise of the outburst (black symbols in Fig. \ref{fig:param_vs_rate}). While $\nu_{LF}$ increases with the \textit{NICER} count rate both in the decay and the reflare of the outburst, the increase is faster during the decay than in the reflare. The fractional rms amplitude evolves differently during the decay of the outburst and the reflare (red and blue symbols in the middle panel of Fig. \ref{fig:param_vs_rate}). In the decay of the outburst the fractional rms remains approximately constant with count rate. During the reflare, the fractional rms appears to decrease slightly with count rate. The FWHM of $L_{LF}$ is anti-correlated with the count rate, while in the decay and the reflare, the FWHM increases with count rate.   

\subsection{Rms-spectra of the type-C QPO}

\begin{figure*}
    \centering
    \includegraphics[width=0.9\textwidth]{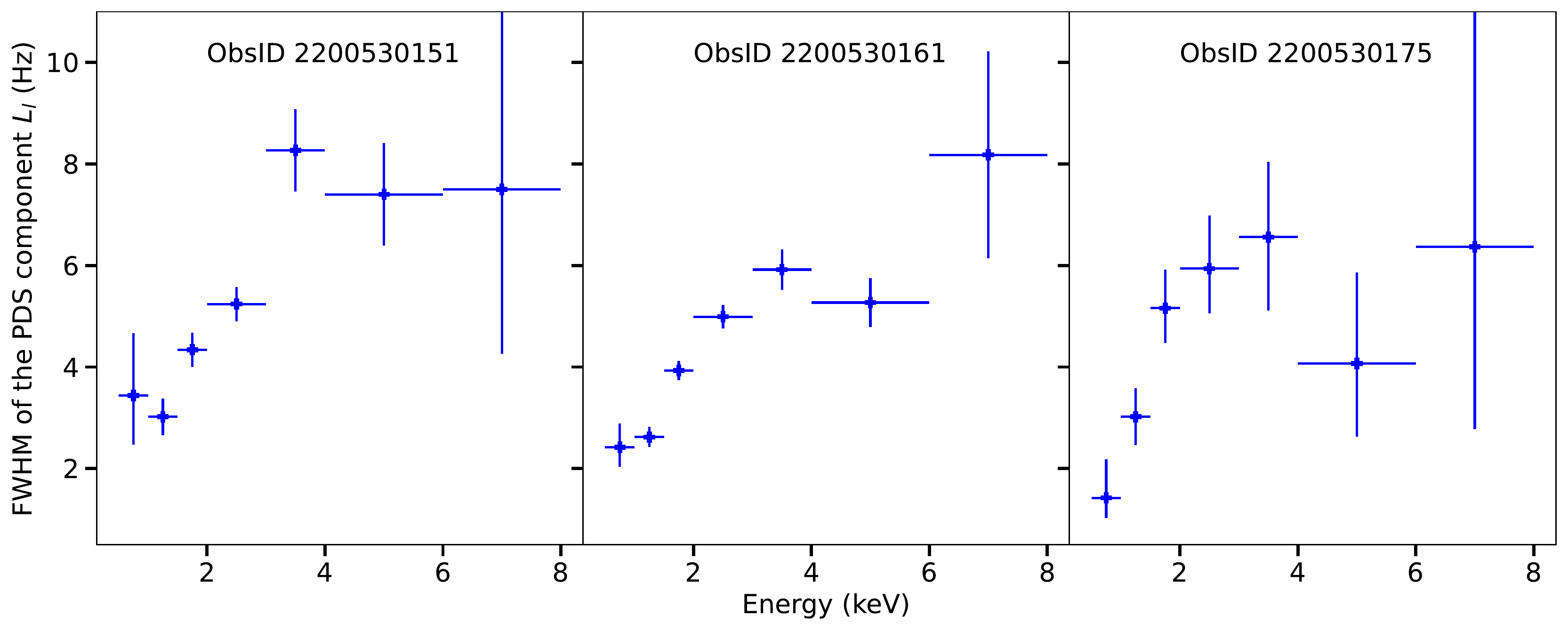}
    \caption{Energy dependence of the high-frequency broad component $L_l$ of the PDS of the reflare.}
    \label{fig:fwhm_vs_energy}
\end{figure*}

\begin{figure*}
    \centering
    \includegraphics[width=0.9\textwidth]{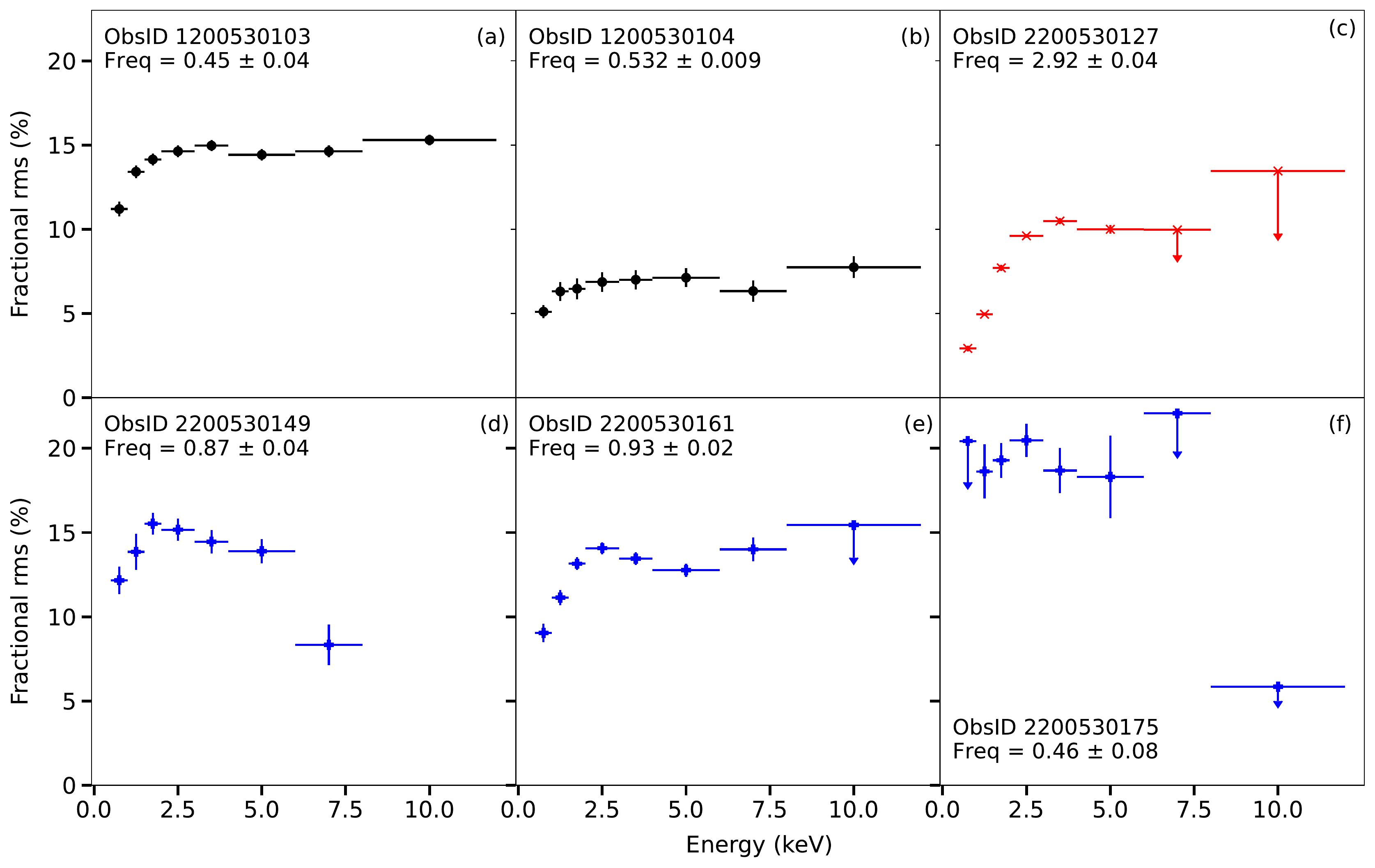}
    \caption{Representative examples of the rms spectra of the type-C QPO in MAXI~J1348--630. Colours and symbols are the same as in Fig. \ref{fig:lc_hardness}. Arrows represent the 95\% confidence upper limits of the fractional rms amplitude. On panel (d), there is an upper limit of 82\% at 10 keV that is not shown due to the selected range for the fractional rms in the y-axis.}
    \label{fig:rms_spec}
\end{figure*}

For each observation showing a type-C QPO, we studied the energy-dependent variability of the QPO. First, we fitted the PDS of the observations with QPOs in the different energy bands as mentioned in Section 2. We found that, during the main outburst, the characteristic frequency and the FWHM of the QPO are always consistent with being constant with energy. The only exception is the component $L_l$, for which the FWHM changes with energy. Fig. \ref{fig:fwhm_vs_energy} shows the FWHM of $L_{l}$ vs. energy. Because the FWHM changes with energy, in order to obtain the rms spectra of the QPO, we fitted the PDS of all the energy bands fixing the FWHM and the centroid frequency of all the components during the main outburst and the reflare to the values corresponding to the full energy band and leaving the FWHM of $L_l$ free. 

We show some representative rms spectra of the type-C QPO of MAXI~J1348--630 in Fig. \ref{fig:rms_spec}. We found different behaviours depending on the moment along the outburst in which the QPO was present. During the rise of the outburst (panels (a) and (b) of Fig. \ref{fig:rms_spec}), the fractional rms amplitude increases with energy from $\sim$6\% at energies below 1 keV to 9--10\% at 10 keV. At 2--4 keV, the growth stabilises and the fractional rms amplitude remains constant as the energy increases further. During the decay of the outburst (panel (c) of Fig. \ref{fig:rms_spec}), the fractional rms amplitude increases with energy up to 2--4 keV from $\sim$4\% at 1 keV to 8--10\% at 4 keV. Above 4 keV, the fractional rms amplitude remains approximately constant. During the rise of the reflare (MJD 58638-MJD 58643), the fractional rms grows from $\sim$15\% at $\sim$1 keV to 20--25\% up to $\sim$2 keV and then decreases with energy down to $\sim$7\% (panel (d) of Fig. \ref{fig:rms_spec}). From MJD 58644 to MJD 58657, the fractional rms amplitude increases with energy from $\sim$10\% at 1 keV to $\sim$16\% at 2--3 keV. Above this energy, the fractional rms amplitude remains constant as the energy increases further (panel (e) on Fig. \ref{fig:rms_spec}). Finally, from MJD 58659 to MJD 58692 (panel (f) on Fig. \ref{fig:rms_spec}), the fractional rms amplitude is consistent with being constant with energy. In order to assess whether the break at 2--4 keV observed in some cases is significant we fitted these rms-spectra with both a straight and a broken-line. The broken-line provides a much better fit ($\chi_{\nu}^{2} \lesssim 1.5$) than the straight line ($\chi_{\nu}^{2} >> 2$). This suggests that the break is significant and that it occurs at $\sim$1.8 keV during the outburst and the reflare.


\subsection{Phase lags of the type-C QPO}

\begin{figure*}
    \centering
    \includegraphics[width=0.9\textwidth]{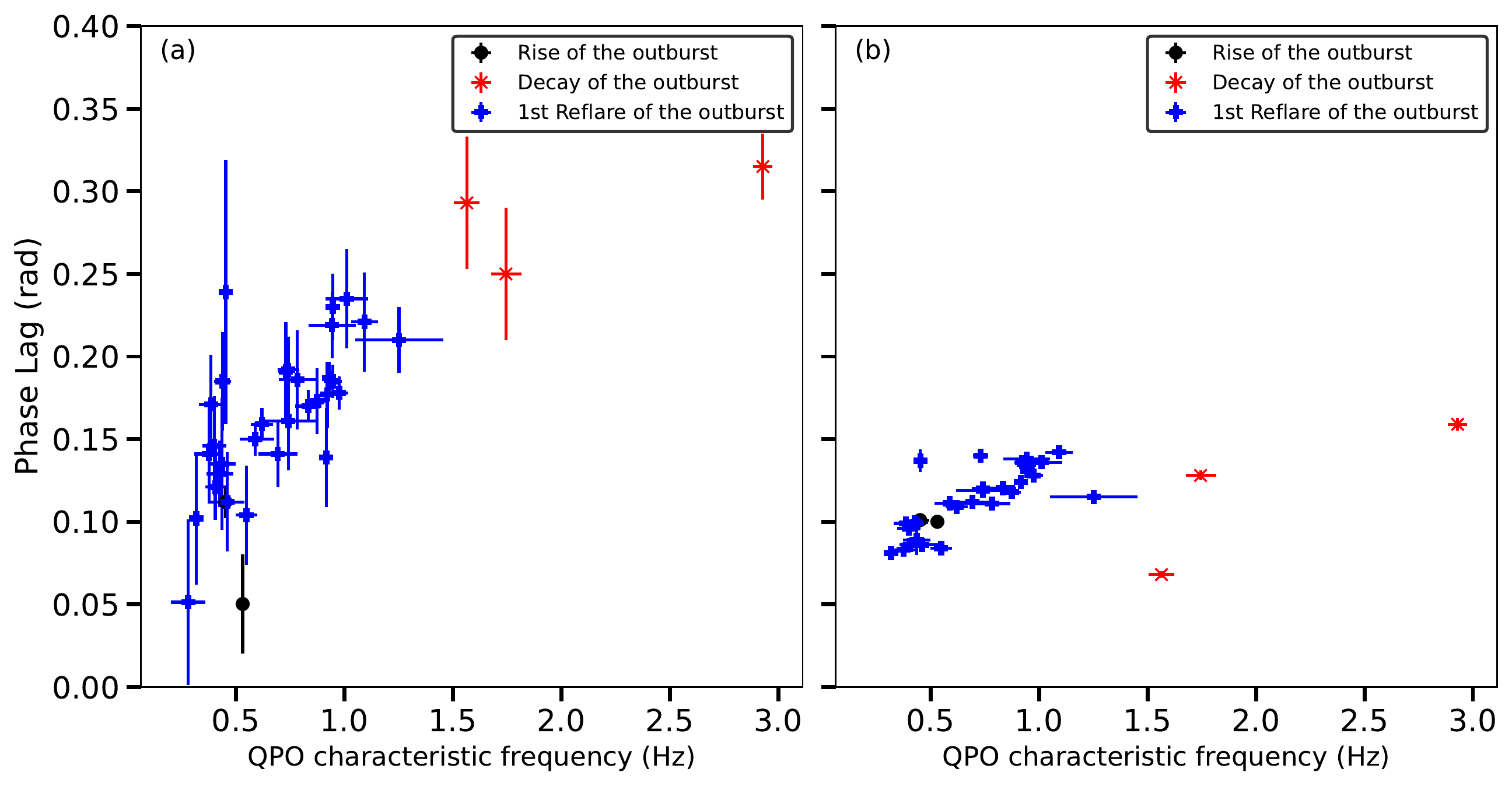}
    \caption{Phase lags as a function of the QPO frequency for the type-C QPO in MAXI~J1348--630. (a): Lags obtained with the new method described in Section 2. (b): Average lags computed within the FWHM around the QPO characteristic frequency. Colours and symbols are the same as in Fig. \ref{fig:lc_hardness}.}
    \label{fig:lag_vs_freq}
\end{figure*}

\begin{figure*}
    \centering
    \includegraphics[width=0.9\textwidth]{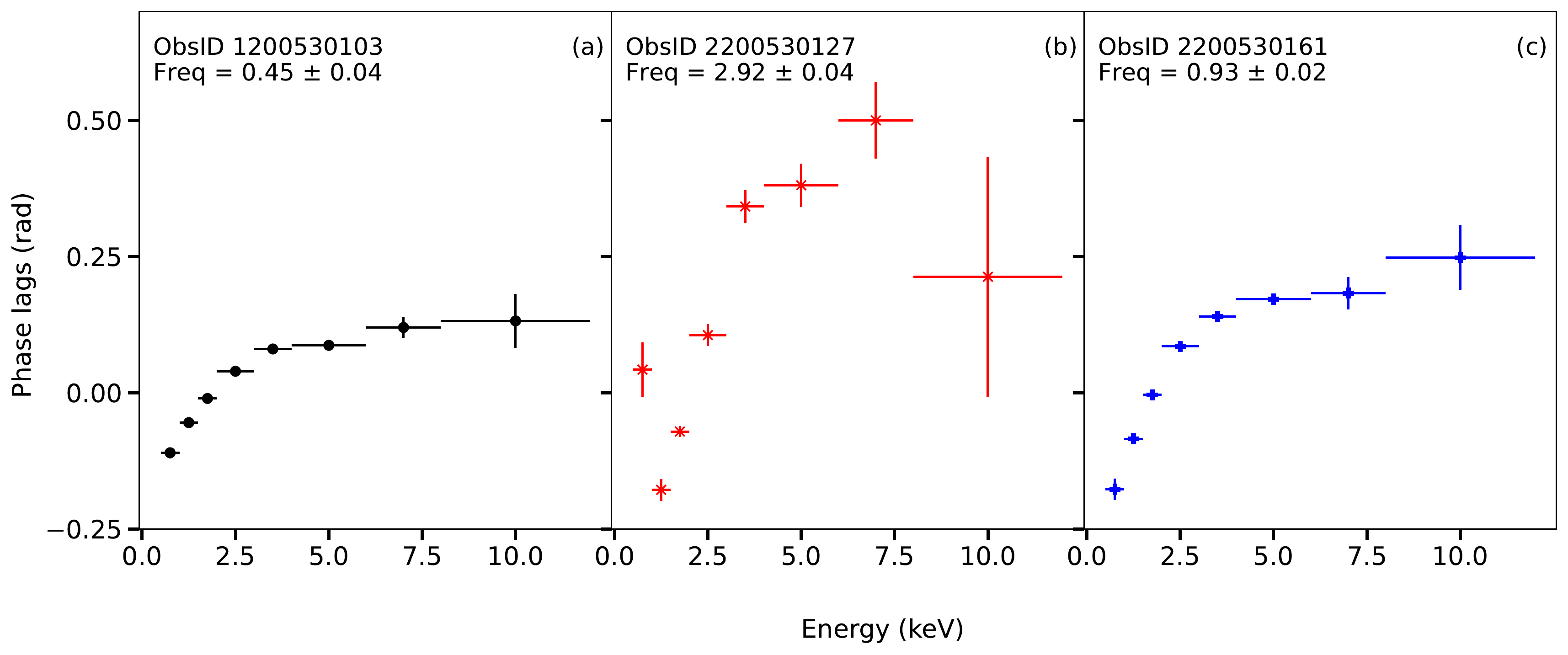}
    \caption{Representative examples of the lag spectra of the type-C QPO in MAXI~J1348--630. Colours and symbols are the same as in Fig. \ref{fig:lc_hardness}.}
    \label{fig:lag_spec}
\end{figure*}

In panel (a) of Fig. \ref{fig:lag_vs_freq} we show the phase lags of the type-C QPO computed between the 0.5--2.0 keV and 2.0--12.0 keV as a function of the characteristic frequency of the QPO, $\nu_{LF}$, following the procedure described in Section 2. During the reflare of MAXI~J1348--630 the QPO phase lags increase with the QPO frequency. While a similar trend may be present during the rise of the outburst, it is difficult to establish this conclusively due to the small number of data points. On the contrary, during the decay of the outburst the phase lags appear to remain constant with the QPO frequency. Assuming that the relation of the phase lags with $\nu_{LF}$ is the same in every phase of the outburst, we observe a break in the slope of the relation when the QPO frequency is $\sim$1--1.5 keV.

In order to compare them with those obtained with the new method, on panel (b) of Fig. \ref{fig:lag_vs_freq} we show the phase lags computed using the traditional method of averaging the cross spectra within one FWHM of the QPO frequency for comparison. From that panel we see that during the rise of the outburst the traditional lags remain more or less constant with the QPO frequency but, as for the lags obtained through the new method, we cannot state this categorically due to the small number of data points. During the decay of the outburst the traditional phase lags increase with the QPO frequency and are lower than the ones obtained with the new method. Finally, during the reflare of the outburst the traditional phase lags increase with the QPO frequency, although over a narrower range than those obtained with the new method.

From a comparison between the two panels in Fig. \ref{fig:lag_vs_freq} it is also apparent that the errors of the traditional phase lags are significantly smaller than those obtained using the new method. The reason is that the signal-to-noise ratio of a lag measurement is proportional to the ratio of the signal of the power to that of the noise in the power spectrum \citep[][]{Bendat10}. In the traditional method the signal power is the combination of the QPO power plus the power of all the other components in the power spectrum over the frequency range used to measure the lags of the QPO, whereas in our method the signal power is only the power of the QPO itself. At the same time, in the traditional method the noise is due to the Poisson component, whereas in our method the ``noise’’ is the sum of the power of all the components in the power spectrum, except the QPO, over the frequency range of the QPO, which is in general higher than the Poisson noise.

In summary, although the values and the errors of the lags obtained via the two methods differ (for the reasons just explained), the general trends with QPO frequency are similar in all phases of the outburst. The differences observed in the lags of the two methods during the decay of the outburst could be due to the small errors of the traditional lags. Since, as we explained before, the measurements and errors of the lags of the QPO using the traditional method are severely affected by the properties of the underlying broadband components in the power spectrum and, therefore, less reliable than those obtained with the new method, in the rest of the paper we only discuss the lags obtained through the new method.



We also studied the lag spectrum for each observation with a type-C QPO. We obtained the lag spectrum following the procedure described in Section 2. Fig. \ref{fig:lag_spec} shows some representative lag spectra of the type-C QPOs of MAXI~J1348--630. In general, the lags of the QPO are hard during the whole outburst and the reflare. During the rise of the outburst the lags increase with energy up to $\sim$3 keV and then remain approximately constant (panel (a) of Fig. \ref{fig:lag_spec}). During the decay of the outburst, in ObsIDs 2200530127 and 2200530129, the lags decrease from 0.5 keV to 1.0 keV and then increase with energy (panel (b) of Fig. \ref{fig:lag_spec}). We cannot rule out that a similar decrease from 0.5 keV to 1.0 keV happens in ObsID 2200530128, also corresponding to the decay of the outburst, due to the large error bars of the lags. During the reflare, the lags increase with energy up to 3--4 keV and then remain approximately constant (panel (c) of Fig. \ref{fig:lag_spec}). In some observations the lags appear to be consistent with zero within errors.

\section{Discussion}

We studied the properties of the type-C QPO of MAXI~J1348--630 during the 2019 outburst and its subsequent reflare. This is the first time that the evolution of the properties of a QPO are studied in the reflare of a low-mass X-ray binary. We detected the type-C QPO in the LHS during the rise and the reflare of the outburst, and during the HIMS in the decay of the outburst. The type-C QPO shows characteristic frequencies from $\sim$0.3 Hz to $\sim$2.9 Hz and fractional rms amplitude from $\sim$6\% to $\sim$30\%. We found that the frequency of the QPO is correlated with the frequencies of the components $L_{b}$, $L_{u}$ and $L_{}$.
%
We studied the energy-dependent fractional rms amplitude and phase lags associated with the type-C QPO. We found that, in most of the observations, the fractional rms amplitude of the QPO increases with energy up to $\sim$1.8 keV and is consistent with being constant above this energy. However, there are two exceptions. At the beginning of the reflare the rms increases up to 1--2 keV and above this energy it remains approximately flat or decreases with energy. During the decay of the reflare, the fractional rms remains approximately constant with energy. We found that the lags of the type-C QPO are hard during the whole outburst and the reflare, except for a few observations in which the lags are consistent with zero. Finally, we also found that the FWHM of the component $L_l$ changes with energy during the reflare.    

\subsection{Evolution of the parameters of the QPO during the reflare}

It is interesting to compare the evolution of the parameters of the QPO during the reflare of MAXI~J1348--630 with those of type-C QPOs observed in other sources during outburst. It is important to point out that most of the previous studies of QPOs were done with data from the Proportional Counter Array (PCA) in \textit{Rossi X-ray Timing Explorer} \citep[\textit{RXTE}, ][]{Bradt93}, covering the energy range 3--25 keV, while here we used \textit{NICER} observations in the 0.5--12 keV energy band. \textit{NICER} data are more affected by the interstellar absorption than those of \textit{RXTE}, which has an effect on the hardness ratios we obtained. Moreover, the contribution of the disc component may also affect the amplitude of the variability we detect \citep[][]{Uttley11}. However, the contribution of the disc component to the total flux in the observations showing a type-C QPO during the reflare is negligible, much lower than the contribution of the corona \citep[][]{Zhang20}. Therefore, we neglect the difference in the energy range since it will not affect our conclusions significantly.

The frequency of the type-C QPO, $\nu_{LF}$ ranges from $\sim$0.29 Hz to $\sim$1 Hz during the reflare, the fractional rms amplitude between $\sim$10\% and $\sim$26\% and the FWHM from $\sim$0.1 Hz to $\sim$0.9 Hz. The values of the parameters of the type-C QPO during the reflare are consistent with those observed in other BH systems during outburst. The frequency of type-C QPOs in other systems ranges between a few mHz and $\sim$30 Hz (e.g., \citealp[ XTE~J1550--564: ][]{Cui99, Sobczak00, Belloni02a}, \citealp[ XTE~J1859+226: ][]{Casella04}, \citealp[GX~339--4: ][]{Belloni05, Motta11} and \citealp[ GRO~1650--40: ][]{Motta12}), while the type-C QPOs in other BH systems are characterised by a fractional rms amplitude of up to $\sim$20\% \citep[e.g., ][]{Sobczak00, Remillard02, Belloni02a, Casella04, Casella05, Belloni05, Motta11, Motta15, Motta16}.  
%
%
Regarding the width of the QPO, in previous works it was studied using the $Q$ factor (see Section 2 for its definition). In general, type-C QPOs show a $Q>10$ \citep[e.g., ][]{Belloni02a, Casella04, Motta11}. Although we studied the width of the QPO using the FWHM, a quick conversion to the $Q$ factor can be done in order to compare the values. The $Q$ factor of the QPO in the reflare of MAXI~~J1348--630 ranges between $\sim$0.4 and $\sim$6.7, which is much lower than what was obtained in other sources. 

Figs. \ref{fig:param_vs_hardness} and \ref{fig:param_vs_rate} show the relations between the parameters of the QPO and the hardness ratio and count rate, respectively. Focusing on the observations during the reflare (blue symbols in both figures), we see that $\nu_{LF}$ and the FWHM of the QPO decrease as the hardness ratio increases, while the fractional rms amplitude increases slightly with hardness ratio. The behaviour of the parameters with the count rate is just the opposite: the frequency and the FWHM increase while the fractional rms amplitude slightly decreases as count rate increases. A similar behaviour of $\nu_{LF}$ with count rate was found in other BH systems: e.g.,  GRS~1915$+$105 \citep{Trudolyubov99a, Reig00}, GRS~1739--278 \citep{Wijnands01}, XTE~J1859$+$226 \citep[][]{Casella05b}, XTE~J1650--500 \citep[][]{Xiao18}, XTE~J1550--564 \citep[][]{Heil11}, Swift~J1842.5--1124 \citep[][]{Zhao16} and EXO~1846--031 \citep[][]{Liu20}. The QPO frequency of GRS~1739--278, on the contrary, increases with the hardness ratio \citep{Wijnands01}. The fractional rms amplitude of the type-C QPOs has also been observed to decrease in  GRS~1915$+$105 \citep{Trudolyubov99a, Reig00}. The width of the QPO, on the other hand, also decreases in Swift~J1842.5--1124 \citep[][]{Zhao16}.  

In summary, despite the difference between the width of the type-C QPO in the reflare of MAXI~J1348--630 and that in other sources, the similarity between the ranges of frequency and fractional rms of the type-C QPOs and the evolution of the frequency with the hardness ratio and count rate suggests that the dynamical mechanism behind the type-C QPOs during the reflare of MAXI~J1348--630 is similar to that behind the type-C QPOs in outbursts of BH LMXBs.    

\subsection{The PBK and the WK relations}

Fig. \ref{fig:freq_corr} shows the relation of $\nu_{LF}$ with $\nu_{b}$ (left panel), $\nu_u$ (middle panel) and $\nu_{l}$ (right panel), and the PBK \citep[][]{Psaltis99} and WK \citep[][]{Wijnands99a} relations (grey symbols in the three panels). Originally, the PBK relation was found between low-frequency kilo-Hertz QPOs and high-frequency kilo-Hertz QPOs of high luminosity NS systems \citep[e.g., ][]{Psaltis98}. \citet{Psaltis99} and \citet{Nowak00} extended this relation to the frequency of the low-frequency QPOs and two broad frequency Lorentzians at high frequencies for low luminosity NS and BH LMXBs. The PBK and WK relations have been observed in several studies \citep[e.g., ][]{Nowak00, Kalemci01, Kalemci03, Belloni02a, Klein-Wolt04, Klein-Wolt08}. We found that $\nu_{LF}$ is correlated with $\nu_b$, $\nu_{u}$ and $\nu_{l}$. While the  $\nu_{LF}$--$\nu_{b}$ relation lies slightly below the WK relation, the $\nu_{LF}$-$\nu_{u}$ and $\nu_{LF}$-$\nu_{l}$ relations lie slightly below and above the PBK relation, respectively. One reason why these relations deviate slightly from the PBK and the WK relations could be that our results are obtained with \textit{NICER}, whereas the PBK and WK relations were obtained using \textit{RXTE} data.
%
The fact that the frequency of the QPO is correlated with the frequencies of $L_b$, $L_u$ and $L_l$ suggests that the dynamical mechanism responsible for all these components is related to the same physical property.

\subsection{Energy dependence of the FWHM of $L_l$}

We found that the FWHM of the component $L_l$ in the reflare changes with energy (Fig. \ref{fig:fwhm_vs_energy}). Since this component is a zero-centred lorentzian, the FWHM is related to the characteristic frequency by $\nu_{max} = FWHM/2$.

There are other BH~LMXBs in which the frequency of a component in the PDS changes with energy, namely GRS~1915+105 \citep{Qu10, Yan12, vandenEijnden17, Yan18}, H~1743--322 \citep{Li13, vandenEijnden17}, XTE~J1550--564 and XTE~J1859+226 \citep[][]{vandenEijnden17} and MAXI~1535--571 \citep[][]{Huang18}.
The behaviour of $L_l$ is similar to the one observed for the QPO of MAXI~J1535--571 with frequencies above 3 Hz presented in \citet{Huang18}. These authors found that the frequency of the QPO increases up to 3--4 keV and above that energy the frequency increases very slowly. A similar behaviour was observed in observations of some of the classes of GRS~1915+105 \citep[][]{Qu10, Yan12, Yan18}. 

Different models have been proposed to explain the energy dependence of the QPO frequency: the global disc oscillation model \citep[GDO, ][]{Titarchuk00}, the radial and orbital oscillation model \citep[ROOM, ][]{Nowak93}, the drift blob model \citep[DBM, ][]{Hua97} and the differential precession of the inner accretion flow \citep{vandenEijnden16}. According to the latter, the QPO is produced at different regions of the inner accretion flow (or corona). The QPO produced at the inner regions of the corona would have a higher frequency than the QPO produced at the outer parts of the corona. This mechanism produces the changes of the QPO frequency with energy. Although the component of MAXI~J1348--630 showing an energy-dependent frequency is formally not a QPO, but a broadband component, we show in the right panel of Fig. \ref{fig:freq_corr} that the frequencies of the type-C QPOs and $L_l$ are correlated. This correlation suggests a similar origin for both frequencies. Therefore, the energy dependence of the FWHM of $L_l$ could be due to the origin of this PDS component at different regions of the corona.
%

\subsection{Rms spectra of the type-C QPOs of MAXI~J1348--630}

Apart from the evolution of the QPO parameters, we also studied the energy dependence of the fractional rms amplitude of the type-C QPO of MAXI~J1348--630. Thanks to \textit{NICER} capabilities, we could extend the study of the rms spectra to energies down to 0.5 keV. We found that the rms spectra of the type-C QPO of MAXI~J1348--630 increases with energy up to $\sim$1.8 keV and above this energy the rms spectra remains approximately constant.

The fact that the fractional rms amplitude is higher at high than at low energies is consistent with what we know from other BH LMXBs: XTE~J1550--564 \citep[e.g, ][]{Kalemci01, Rodriguez04, Sobolewska06, Heil11, Zhang17b}, GRS~1915$+$105 \citep[e.g., ][]{Reig00, Tomsick01c, Rodriguez04a, Zhang20b, Karpouzas21}, MAXI~1659--152 \citep[][]{Kalamkar15}  XTE~J1859$+$226 \citep[e.g., ][]{Sobolewska06, Casella04, Casella05b}, MAXI~J16131--479 \citep[][]{Bu21}, XTE~J1650--500 \citep[][]{Chatterjee20a}, GX~339--4 \citep[e.g., ][]{Sobolewska06, Motta11, Zhang20b} and MAXI~J1535--571 \citep[][]{Huang18}, among others. However, there is an important difference between the previous sources and MAXI~J1348--630. While the fractional rms of type-C QPOs in those other systems increases up to $\sim$10 keV and above this energy it remains approximately constant, in the case of  MAXI~J1348--630, as we mentioned before, the energy from which the fractional rms of the QPO remains constant is $\sim$1.8 keV.

From \citet{Zhang20} we know that the total emission of MAXI~J1348--348 was dominated by the Comptonised component during the observations showing type-C QPOs, suggesting that this component is responsible to the variability of the system rather than the disc component and, therefore, of the energy dependence of the fractional rms amplitude and the phase lags. This scenario has been studied previously for QPOs \citep[e.g., ][]{Lee98, Lee01, Mendez01, Kumar14, Karpouzas20, Zhang20b, Karpouzas21, Garcia22} and the broadband noise \citep[e.g., ][]{Gierlinski05, Alabarta20}. The fractional rms amplitude is higher at higher energies, where the emission is dominated by the Comptonised component. On the contrary, at lower energies, where the contribution of the disc is higher, the fractional rms amplitude is lower. This scenario agrees with what we found for the rms spectra of the type-C QPO in MAXI~1348--630. 

\subsection{Phase lags of the type-C QPOs of MAXI~J1348--630}

We also studied the frequency and energy dependence of the phase lags associated with the type-C QPO of MAXI~J1348--630. As we explained in Section 2, we used a new technique to obtain the phase lags in the different energy bands. This technique allows us to remove the contribution of other components of the PDS to the lags of the QPO, and ensures that we measure the lags of the QPO itself. 

Regarding the frequency dependence, we found that the phase lags of the QPO are always positive above 0.5 Hz and the lags appear to increase with QPO frequency during the rise and the reflare, as we show in panel (a) of Fig. \ref{fig:lag_vs_freq}. On the contrary, the QPO lags are consistent with being constant with QPO frequency during the decay of the outburst. The different relation between the lags of the QPO and the QPO frequency at the decay compared with the rise and the reflare suggests that the geometry of the corona is different in the rise and reflare compared to the decay of the outburst. Fig. \ref{fig:hid} shows that the observations corresponding to the decay of the outburst are softer than those at the rise and the reflare. The presence of a strong disc component in the observations showing the type-C QPO during the decay could be the reason for this different behaviour. 

During the reflare and the rise of the outburst, the phase lags computed using the traditional method, within one FWHM around the QPO frequency (panel (b) of Fig. \ref{fig:lag_vs_freq}), behave in a similar way as the phase lags obtained with the new method described in Section 2. During the decay of the outburst, however, the phase lags computed in the traditional way increase with the QPO frequency whereas those obtained with the new method remain constant. Despite the similar results, as we described in Section 2, the traditional method includes the contribution of other Lorentzian components than the QPO, which can modify the lags. In our case, the difference is more clear in the decay of the outburst, where the lags obtained with the traditional method are lower than those we obtain with the new method.

The lags vs. QPO frequency relations that we show in Fig. \ref{fig:lag_vs_freq} (both in panels (a) and (b)) are different from those observed in other sources. For instance, in GRS 1915+105 the QPO lags are hard at low QPO frequency, decrease with QPO frequency, become zero at a QPO frequency of $\sim$2 Hz, and turn soft above that frequency \citep[][]{Reig00, Qu10, Pahari13, Zhang20b}. On the contrary, the lags of the type-C QPO in MAXI~J1348--630 do not decrease with frequency and there is no change from hard to soft lags. The lags of the QPO in other black-hole candidates are consistent with zero up to 1--5 Hz, and then increase or decrease with frequency, depending on the inclination of the source \citep[][]{vandenEijnden17}. The fact that the lag-frequency relation of the type-C QPO is different for other sources makes this source an interesting target to study the radiative and geometrical origin of the type-C QPOs.

As for the energy dependence of the QPO lags, we found that, in general, the phase lags of the QPO are hard during the outburst and the reflare. There are only a few ObsIDs in which the QPO lags are consistent with zero within errors. Hard lags associated with type-C QPOs have been observed in other BH systems: GX~339--4 \citep[][]{Zhang17b}, GRS~1915$+$105 \citep[at low QPO frequencies, ][]{Cui99, Pahari13, Zhang20b, Karpouzas21, Garcia22}, MAXI~J1820$+$070 \citep[][]{Mudambi20, Wang20} and XTE~J1650--500 \citep[][]{Chatterjee20a}. On the contrary, some sources show soft lags associated to type-C QPOs: GRS~1915$+$105 \citep[at high QPO frequencies, ][]{Reig00, Pahari13, vandenEijnden17, Zhang20b}, XTE~J1859$+$226, XTE~J1550--564, MAXI~J1659--152, H1743--322 and MAXI~J1543--564 \citep[][]{vandenEijnden17}.

Despite the fact that the QPO lags are hard during the rise and the decay of the outburst and the reflare, there are differences in the shape of the rms spectra between the different phases of the outburst. As we can see in panel (a) of Fig. \ref{fig:lag_spec}, during the rise of the outburst, the lags increase from 0.5 keV up to 2--3 keV, and are consistent with being constant or slightly decrease above this energy. During two observations of the decay (ObsIDs 2200530127 and 2200530129), the lags of the QPO decrease from 0.5 keV to 1 keV, and above this energy the QPO lags increase. During ObsID 2200530128, also corresponding to the decay of the outburst, the error bars of the QPO lags below 1 keV do not allow us to tell whether the lags show the same trend. Finally, during the reflare, the lags of the QPO increase with energy up to 2--3 keV and above that energy they are consistent with being constant. The different QPO lag spectra during the rise, the decay and the reflare reinforces the conclusion obtained from the lag vs. QPO frequency relation that the geometry of the region responsible for the radiative origin of the QPO is different in these three phases of the outburst.

Different models have been proposed to explain the shape of the energy dependence of phase lags and the detection of both hard and soft lags. One model is the Comptonisation model presented in \citet{Nobili00}. This model consists of a corona with two components: an inner hot and optically thick component and an optically thin component. \citet{Nobili00} also assumes that the accretion disc is truncated at a radius $R_{in}$. When the truncation radius is large, the corona up-scatters the soft photons coming from the accretion disc. As a consequence, the hard photons come later than the soft photons and the lags are hard. On the contrary, when the inner disc radius is small, the inner part of the accretion disc lies inside the hot part of the corona, producing a down-scattering of soft photons that leads to soft lags. In this model, what determines whether the lags are hard or soft are the changes on the inner radius of the accretion disc. Since the phase lags associated with the type-C QPOs of MAXI~J1348--630 are hard, our results are in agreement with the model of \citet{Nobili00} only if the inner radius of the accretion disc is truncated. Considering that the type-C QPOs were found in observations corresponding to the LHS and the HIMS, it may be possible that the inner disc radius is truncated far enough to not produce the down-scattering of the soft photons that accounts for the soft lags in this model. 

Although the model of \citet{Nobili00} can explain the phase lags observed in MAXI~J1348--630, the study of the energy dependence of the fractional rms amplitude of type-C QPOs points out to a coronal origin for type-C QPOs, rather than an accretion disc origin \citep[e.g., ][]{Casella04, Sobolewska06, Motta11}, as we discussed in section 4.4. We can also explain the phase lags from the Comptonisation model presented in \citet{Karpouzas20}, which is based on the model of \citet{Lee01} and \citet{Kumar14}. In this model, the corona up-scatters the soft photons coming from the accretion disc, producing a delay of the hard photons. In addition to this process, this model takes account feedback between the corona and the accretion disc. Some of the Comptonised photons in the corona impinge back onto the disc and heat the accretion disc that emits at lower energies than the corona. This process produces a time delay of soft photons that translates into soft lags. As explained in this section and in Section 3, we only found hard lags for the type-C QPO of MAXI~J1348--630. Because of that, the shape of the energy dependence of the phase lags of this system is in agreement with the Comptonisation process described in \citet{Karpouzas20} if the system shows a very small feedback. Connecting this interpretations with that from the model of \citet{Nobili00}, if the disc is truncated far enough for a given corona size \citep[similar to][for GRS~1915+105]{Karpouzas21} then feedback is indeed expected to be low, leading to hard rather than soft lags.

\subsection{Inclination of the disc of MAXI~J1348--630}

MAXI~J1348--630 was first identified as a low-inclination system based on reflection studies of \textit{NuSTAR} observations \citep[][]{Anczarski20, Chakraborty21}, that found values corresponding to low-inclination systems: $\sim$28$^{\circ}$ and values in the range 30$^{\circ}$--40$^{\circ}$, respectively. Alternatively, \citet{Carotenuto22} obtained an inclination of $\sim$29$^{\circ}$ for the jet of the system with respect to the line of sight. From this value, we can explore two scenarios in order to estimate the inclination of the accretion disc. In the first one, both the disc and the jet are perpendicular, and the inclination of the disc would be $\sim$29$^{\circ}$, the same as the inclination angle of the jet. In the second scenario, the disc and the jet are not perpendicular, in which case it is impossible to estimate the inclination of the disc from that of the jet.

As we mentioned in Section 1, a potential relation between the fractional rms amplitude and the phase lags of type-C QPOs was found by \citet{Motta15} and \citet{vandenEijnden17}. Based on these studies, we can check whether the properties of the type-C QPO of MAXI~J1348--630 are consistent with a low- or a high-inclination system.

\citet{Motta15} found evidence that high-inclination sources show QPOs with higher fractional rms amplitude than low-inclination sources. During the reflare, the QPO of MAXI~J1348--630 shows frequencies below 1 Hz and fractional rms amplitudes higher than 10\%. According to the results of \citet{Motta15}, MAXI~J1348--630 could be a high-inclination source, which is in disagreement with our estimation of the inclination of the disc based in \citet{Carotenuto22}. However, the fractional rms of the QPO during the decay of the outburst (with frequencies between 1 Hz and 3 Hz) is $\sim$6\%, which is more consistent with the low-inclination system population presented in \citet{Motta15}. We cannot, therefore, estimate whether MAXI~J1348--630 is a low-inclination system from the amplitude of the type-C QPO.

\citet{vandenEijnden17} suggested that the phase lags of type-C QPOs depend on the inclination of the source. At QPO frequencies higher than 2 Hz, low-inclination sources show hard lags while high-inclination sources show soft lags. On the contrary, at low QPO frequencies, all sources display hard lags. As we show in Fig. \ref{fig:lag_spec}, the type-C QPO of MAXI~J1348-–630 shows in almost all the observations hard lags, both below and above 2 Hz. The proposal of \citet{vandenEijnden17} suggests that this is a low-inclination system, in agreement with \citet{Anczarski20} and \citet{Chakraborty21}. Nevertheless, we should be cautious about this conclusion since we only have one observation with a QPO above 2 Hz and the method to obtain the phase lags is different from that used in \citet{vandenEijnden17}.

Apart from a potential relation between the phase lags and the inclination of the system, \citet{vandenEijnden17} also showed an energy dependence of the QPO frequency for three high-inclination systems (GRS~1915+105, H~1743--322 and XTE~J1550--564) and XTE~J1859+226, the latter with an unknown inclination. These authors found that the QPO frequency changes with energy above 6--7 Hz. Below that frequency, the QPO frequency remains constant as the energy increases further. The QPO frequency of the type-C QPO of MAXI~J1348--630 remains constant with energy which, in agreement with the conclusion obtained from the phase lags, suggests that MAXI~J1348--630 is a low-inclination system. However, the frequency of the QPO of the four systems presented by \citet{vandenEijnden17} started to vary at frequencies above 6--7 Hz, above the maximum frequency of the type-C QPO of MAXI~J1348--630. Because of that, we cannot safely conclude that MAXI~J1348--630 is a low-inclination system.

\section{Summary and conclusions}

In this work we present the study of the properties of the type-C QPO during the 2019 outburst and reflare of MAXI~J1348--630. This is the first study of the evolution of the properties of a QPO during a reflare of a BH~LMXB. The summary of our conclusions are the following:

\begin{enumerate}
    \item The properties of the type-C QPO during the reflare are similar to those of the type-C QPO during the main outburst of the system and the type-C QPOs observed during the main outburst of other sources. This result reinforces the idea that both outbursts and reflares are driven by the same physical processes.
    \item The frequency of the type-C QPO is correlated with the frequencies of the components $L_{u}$, $L_{b}$ and $L_{l}$, but these correlations lie slightly down the PBK and WK relations. These correlation suggests that the dynamical mechanisms responsible for these components' frequencies are related to the same physical process.
    \item The characteristic frequency of the component $L_{l}$ varies with energy, as has been observed in QPOs in other systems.
    \item The relation between the QPO phase lags and the QPO frequency suggests that there are differences in the geometry of the system at the different phases of the outburst. 
    \item The fractional rms amplitude of the type-C QPO increases with energy up to $\sim$1.8 keV and then is consistent with being constant with energy. The phase lags of the type-C QPO are hard during the main outburst and the reflare. Both the energy-dependent fractional rms amplitude and the phase lag-spectra can be explained in terms of Comptonisation. 
    \item Our estimation of the inclination angle of the accretion disc, the hard lags of the type-C QPO and the lack of energy dependence of the QPO frequency point to a low-inclination system in agreement with previous estimations of the inclination. However, the lack of QPOs at frequencies higher than $\sim$2.9 Hz makes impossible to estimate the inclination of the system with certainty.
\end{enumerate}

\begingroup
\setlength{\tabcolsep}{14pt} 
\renewcommand{\arraystretch}{1.7} 
\begin{table*}
\tiny
\resizebox{\textwidth}{!}{
\begin{tabular}{cccccc}
\hline 
\textbf{ObsID} & \textbf{MJD} & \textbf{QPO frequency (Hz)} & \textbf{QPO FWHM (Hz)} & \textbf{QPO Fractional rms amplitude (\%)} & \textbf{Part of the outburst} \\
\hline
1200530103 & 58511 & $0.45\pm0.04$ & $0.29\pm0.10$ & $13.5\pm1.9$ & Rise \\
\vspace{\fill}
1200530104 & 58512 & $0.532\pm0.009$  & $0.08\pm0.02$ & $6.46\pm0.77$ & Rise \\
2200530127 & 58603 & $2.93\pm0.04$ & $1.49\pm0.17$ & $6.14\pm0.33$ & Decay \\
2200530128 & 58605 & $1.74\pm0.07$ & $1.01\pm0.31$ & $5.65\pm_{0.14}^{0.09}$ & Decay \\
2200530129 & 58606 & $1.56\pm0.06$ & $0.86\pm0.25$ & $5.94\pm0.84$ & Decay \\
2200530145 & 58638 & $0.62\pm0.05$ & $1.16\pm0.06$ & $25.47\pm1.96$ & Reflare \\
2200530146 & 58639 & $0.69\pm0.09$ & $0.99\pm0.16$ & $22.13\pm2.25$ & Reflare \\
2200530147 & 58640 & $0.77\pm0.20$ & $1.12\pm0.22$ & $17.94\pm2.79$ & Reflare \\
2200530148 & 58641 & $0.78\pm0.09$ & $0.72\pm0.21$ & $14.89\pm3.02$ & Reflare \\
2200530149 & 58641 & $0.87\pm0.04$ & $0.83\pm0.10$ & $14.38\pm1.04$ & Reflare \\
2200530150 & 58643 & $0.83\pm0.06$ & $1.18\pm0.10$ & $21.11\pm1.42$ & Reflare \\
2200530151 & 58644 & $0.95\pm0.05$ & $0.79\pm0.09$ & $15.03\pm0.99$ & Reflare \\
2200530152 & 58645 & $0.94\pm0.11$ & $1.09\pm0.21$ & $15.23\pm1.97$ & Reflare \\
2200530153 & 58646 & $0.95\pm0.03$ & $0.70\pm0.09$ & $13.19\pm0.76$ & Reflare \\
2200530155 & 58649 & $1.09\pm0.06$ & $0.73\pm0.16$ & $11.53\pm1.30$ & Reflare \\
2200530156 & 58650 & $0.96\pm0.04$ & $0.97\pm0.18$ & $11.60\pm2.31$ & Reflare \\
2200530157 & 58651 & $1.01\pm0.09$ & $0.84\pm0.25$ & $11.79\pm1.69$ & Reflare \\
2200530159 & 58653 & $0.92\pm0.03$ & $0.44\pm0.09$ & $10.39\pm0.96$ & Reflare \\
2200530160 & 58654 & $0.92\pm0.03$ & $0.61\pm0.11$ & $12.56\pm1.19$ & Reflare \\
2200530161 & 58655 & $0.93\pm0.02$ & $0.54\pm0.06$ & $12.68\pm0.39$ & Reflare \\
2200530162 & 58656 & $0.98\pm0.04$ & $0.70\pm0.11$ & $13.15\pm0.76$ & Reflare \\
2200530163 & 58657 & $0.74\pm0.05$ & $0.60\pm0.10$ & $14.79\pm1.69$ & Reflare \\
2200530164 & 58659 & $0.59\pm0.08$ & $0.90\pm0.11$ & $26.83\pm_{3.78}^{5.79}$ & Reflare \\
2200530165 & 58660 & $0.73\pm0.04$ & $0.46\pm0.13$ & $12.21\pm1.64$ & Reflare \\
2200530166 & 58660 & $0.74\pm0.13$ & $0.58\pm0.22$ & $13.63\pm_{2.19}^{3.67}$ & Reflare \\
2200530169 & 58672 & $0.55\pm0.05$ & $0.42\pm0.09$ & $14.66\pm2.38$ & Reflare \\
2200530171 & 58677 & $0.44\pm0.06$ & $0.39\pm0.15$ & $14.76\pm2.03$ & Reflare \\
2200530172 & 58678 & $0.45\pm0.02$ & $0.11\pm0.05$ & $10.39\pm1.92$ & Reflare \\
2200530174 & 58680 & $0.43\pm0.06$ & $0.62\pm0.08$ & $20.59\pm2.43$ & Reflare \\
2200530175 & 58681 & $0.46\pm0.08$ & $0.57\pm0.11$ & $19.15\pm1.83$ & Reflare \\
2200530176 & 58682 & $0.41\pm0.05$ & $0.47\pm0.08$ & $17.15\pm2.04$ & Reflare \\
2200530180 & 58686 & $0.40\pm0.05$ & $0.48\pm0.09$ & $16.61\pm2.71$ & Reflare \\
2200530182 & 58688 & $0.39\pm0.06$ & $0.40\pm0.09$ & $15.83\pm3.16$ & Reflare \\
2200530183 & 58689 & $0.44\pm0.04$ & $0.43\pm0.08$ & $15.19\pm1.97$ & Reflare \\
2200530184 & 58690 & $0.38\pm0.07$ & $0.57\pm0.09$ & $20.66\pm2.42$ & Reflare \\
2200530185 & 58691 & $0.32\pm0.02$ & $0.26\pm0.06$ & $15.87\pm3.15$ & Reflare \\
2200530186 & 58692 & $0.29\pm0.08$ & $0.34\pm0.11$ & $15.26\pm6.55$ & Reflare \\
\hline
\end{tabular}}
\caption{Table with the \textit{NICER} observations analysed in this work.}
\label{Tab:nicer_obsid}
\end{table*}
\endgroup

\section*{Acknowledgements}

This work is based on observations made by the \textit{NICER} X-ray mission supported by NASA and has made use of \textit{RXTE} data. This research has made use of data and software provided by the High Energy Astrophysics Science Archive Research Center (HEASARC), a service of the Astrophysics Science Division at NASA/GSFC and the High Energy Astrophysics Division of the Smithsonian Astrophysical Observatory. The authors acknowledge Dr. P. Saikia and Dr. M.C. Baglio for the discussion about the inclination of the jet of black-hole X-ray binaries. The authors also thank Samuele Crespi for the discussion about the statistical errors of the different methods used to get the phase lags for the type-C QPO. KA acknowledges support from a UGC-UKIERI Phase 3 Thematic Partnership (UGC-UKIERI-2017-18-006; PI: P. Gandhi). MM, FG and KK acknowledge support from the research programme Athena with project number 184.034.002, which is (partly) financed by the Dutch Research Council (NWO). FG is a CONICET researcher. FG also acknowledges support by PIP 0113 (CONICET). This work received financial support from PICT-2017-2865 (ANPCyT). DA acknowledges support from the Royal Society. LZ acknowledges support from the Royal Society Newton Funds.

\section*{Data Availability}

The data underlying this article are publicly available in the High Energy Astrophysics Science Archive Research Center (HEASARC) at \url{https://heasarc.gsfc.nasa.gov/db-perl/W3Browse/w3browse.pl}



\bibliographystyle{mnras}
\bibliography{bib_all.bib} 




\appendix


\bsp	
\label{lastpage}
\end{document}